\documentclass[aps,prd,preprintnumbers,groupedaddress,nofootinbib,amssymb,eqsecnum,notitlepage]{revtex4}
\usepackage{here}
\usepackage[dvipdfmx]{graphicx}
\usepackage{amsmath,amsthm,amssymb}
\usepackage{bm}
\usepackage{color}

\usepackage{amsfonts}
\usepackage{dcolumn}
\usepackage{hyperref}
\allowdisplaybreaks[1]

\begin{document}
\newcommand{\newc}{\newcommand}

\newcommand{\ben}{\begin{eqnarray}}
\newcommand{\een}{\end{eqnarray}}
\newc{\be}{\begin{equation}}
\newc{\ee}{\end{equation}}
\newc{\ba}{\begin{eqnarray}}
\newc{\ea}{\end{eqnarray}}
\newc{\bea}{\begin{eqnarray*}}
\newc{\eea}{\end{eqnarray*}}
\newc{\D}{\partial}
\newc{\ie}{{\it i.e.} }
\newc{\eg}{{\it e.g.} }
\newc{\etc}{{\it etc.} }
\newc{\etal}{{\it et al.}}
\newcommand{\nn}{\nonumber}
\newc{\ra}{\rightarrow}
\newc{\lra}{\leftrightarrow}
\newc{\lsim}{\buildrel{<}\over{\sim}}
\newc{\gsim}{\buildrel{>}\over{\sim}}
\newc{\aP}{\alpha_{\rm P}}

\title{Coupled vector dark energy}

\author{Shintaro Nakamura, Ryotaro Kase, and Shinji Tsujikawa}

\affiliation{Department of Physics, Faculty of Science, 
Tokyo University of Science, 1-3, Kagurazaka,
Shinjuku-ku, Tokyo 162-8601, Japan}

\begin{abstract}
We provide a general framework for studying the evolution of background 
and cosmological perturbations in the presence of a vector field $A_{\mu}$ 
coupled to cold dark matter (CDM).
We consider an interacting Lagrangian of the form $Q f(X) T_c$, where 
$Q$ is a coupling constant, $f$ is an arbitrary function of 
$X=-A_{\mu}A^{\mu}/2$, and $T_c$ is a trace of the CDM 
energy-momentum tensor. The matter coupling affects the no-ghost condition
and sound speed of linear scalar perturbations deep inside the sound horizon, 
while those of tensor and vector 
perturbations are not subject to modifications. 
The existence of interactions also modifies the no-ghost condition 
of CDM density perturbations.
We propose a concrete model of coupled vector dark energy with 
the tensor propagation speed equivalent to that of light.
In comparison to the $Q=0$ case, we show that the decay of CDM 
to the vector field leads to the phantom 
dark energy equation of state $w_{\rm DE}$ closer to $-1$. 
This alleviates the problem of observational incompatibility of  
uncoupled models in which $w_{\rm DE}$ significantly deviates from $-1$. 
The maximum values of $w_{\rm DE}$ reached during the matter era are 
bounded from the CDM no-ghost condition of future de Sitter solutions.

\end{abstract}

\date{\today}


\maketitle

\section{Introduction}
\label{introsec}

The energy density of today's universe is dominated by 
two unknown components-- dark energy and 
dark matter \cite{dreview1,dreview2,dreview3,dreview4,dreview5,dmreview1,dmreview2}.
Dark energy gives rise to the late-time cosmic acceleration through 
an effective negative pressure, 
while dark matter leads to the growth of structures through 
gravitational clusterings. Although the two dark components 
have different characteristics, they can be potentially coupled 
to each other. The existence of such interactions generally 
modifies the cosmic expansion and growth histories, so 
the coupled models can be distinguished from the 
$\Lambda$-cold-dark-matter ($\Lambda$CDM) model 
by exploiting numerous 
observational data including supernovae type Ia (SN Ia) \cite{SN1,SN2}, 
cosmic microwave background (CMB) temperature 
anisotropies \cite{WMAP,Planck13}, baryon acoustic oscillations (BAOs) \cite{BAO}, 
and redshift-space distortions (RSDs) \cite{RSD1,RSD2}.

For a canonical scalar field $\varphi$, Wetterich \cite{Wetterich} first proposed 
a coupled quintessence scenario in which $\varphi$ interacts with 
CDM. In this model, the continuity equation of dark energy
is sourced by the term $\beta \rho_c \dot{\varphi}$, where 
$\beta$ is a coupling constant, $\rho_c$ is the dark matter density, and 
$\dot{\varphi}$ is the time derivative of $\varphi$. 
Such a coupling arises in Brans-Dicke theories \cite{Brans} after a conformal 
transformation to the Einstein frame \cite{FM,fRreview}.
The quintessence field $\varphi$ drives 
the cosmic acceleration in the presence of a shallow potential 
$V(\varphi)$, e.g., the exponential potential 
$V(\varphi)=V_0 e^{-\lambda \varphi/M_{\rm pl}}$ with 
$|\lambda|<{\cal O}(1)$ (where $M_{\rm pl}$ is the reduced Planck 
mass) \cite{quin1,quin2,quin3,quin4,quin5}.
Amendola \cite{Amendola99} showed that there exists a scaling 
$\varphi$-matter-dominated epoch ($\varphi$MDE) 
during which the coupling gives rise to 
a constant dark energy density parameter $\Omega_{\varphi}=2\beta^2/3$.

The coupling $\beta$ in interacting quintessence is constrained to be 
$\beta<0.062$ at 95\,\% CL from the Planck CMB data alone, 
but the joint datasets of CMB, SN Ia, BAOs, 
and today's Hubble constant $H_0$ lead to the marginalized posterior distribution
with a peak around $\beta=0.036$ \cite{Ade15}. 
Thus, the possibility for sizable interactions 
between the two dark components remains in current observations.

There are also dark energy models in which a noncanonical scalar 
field is coupled to dark matter, including 
k-essence \cite{Piazza,Tsuji04,Gumjudpai}, 
Horndeski \cite{Gomes1,Gomes2,ABDG,Frusciante:2018aew}, 
and DHOST theories \cite{Koyama}. 
They are mostly based on the interacting term 
$\beta \rho_c \dot{\varphi}$ in the dark energy continuity equation. 
In such noncanonical theories, it is also possible to realize the 
$\varphi$MDE followed by late-time cosmic 
acceleration \cite{Amendola06,Frusciante:2018aew}.
There are also models with more phenomenological choices of 
couplings between two dark sectors \cite{Dalal:2001dt,Zimdahl:2001ar,Chimento:2003iea,Wei:2006ut,Amendola:2006dg,Guo:2007zk,Wang1,Wang2}, 
e.g., $\beta H \rho_c$, where $H$ is the Hubble 
expansion rate. In the latter approach, it is generally 
difficult to identify corresponding Lagrangians and associated 
stability conditions (e.g., no ghosts) of perturbations.

The scalar field is not only the possibility for realizing late-time 
cosmic acceleration, but a massive vector field can also be the source 
for dark energy. In generalized Proca (GP) theories with a 
vector field $A_{\mu}$ breaking the $U(1)$ gauge symmetry \cite{Heisenberg,Tasinato1,Tasinato2,Fleury,Hull,Allys,Jimenez16,Allys2}, 
the time-dependent temporal component of $A_{\mu}$ can give 
rise to self-accelerating de Sitter attractors preceded by 
a constant phantom dark energy equation of state $w_{\rm DE}$ 
during the matter era \cite{DeFelice:2016yws,DeFelice:2016uil}. 
The dark energy models given by the Lagrangian 
$L=M_{\rm pl}^2R/2-(1/4)F_{\mu \nu}F^{\mu \nu}+b_2X^{p_2}
+b_3X^{p_3} \nabla_{\mu}A^{\mu}$, 
where $R$ is the Ricci scalar, $F_{\mu \nu}=
\partial_{\mu}A_{\nu}-\partial_{\nu}A_{\mu}$ 
is the field strength, and $b_2,b_3,p_2,p_3$ are constants 
with $X=-A_{\mu}A^{\mu}/2$, exhibit a better compatibility with the 
datasets of SN Ia, CMB, BAOs, RSDs, and $H_0$ 
in comparison to the $\Lambda$CDM model \cite{deFelice:2017paw}. 
This property persists even with the integrated-Sachs-Wolfe (ISW)
effect and galaxy cross-correlation data, by reflecting the fact 
that the existence of intrinsic vector modes can generate positive 
cross-correlations \cite{Nakamura:2018oyy}.

The natural question arises as to what happens in the presence of 
couplings between the massive vector field and CDM. 
For this purpose, we introduce the interacting Lagrangian 
of the form $L_{\rm int}=-Q f(X)\rho_c$ in this paper, 
where the vector field is coupled to the 
CDM density $\rho_c$ with an arbitrary coupling $f(X)$. 
We consider cubic-order GP theories with baryons 
and radiations taken into account and obtain the 
background equations of motion on the flat
Friedmann-Lema\^{i}tre-Robertson-Walker (FLRW) spacetime
to study the dynamics of coupled dark energy 
from the radiation era to today (see Refs.~\cite{Wei:2006tn,Wei:2006gv} 
for other works about coupled vector dark energy). 
We compute the second-order actions of tensor, vector, 
and scalar perturbations and derive conditions for the 
absence of their ghost and Laplacian instabilities.

For the power-law coupling $f(X)=(X/M_{\rm pl}^2)^q$ with 
the Lagrangian $L=-(1/4)F_{\mu \nu}F^{\mu \nu}+b_2X
+b_3X^{p_3} \nabla_{\mu}A^{\mu}$ in the vector sector, 
we show that, for $Q<0$, the dark energy equation 
of state can be larger than that in the uncoupled case 
during the matter era.
When $Q=0$ the models with $p_3 \le 2$ are in tension 
with observational datasets 
due to the large deviation of $w_{\rm DE}$ 
from $-1$ \cite{deFelice:2017paw,Nakamura:2018oyy}. 
For example, the vector Galileon ($p_3=1$) gives the 
value $w_{\rm DE}=-2$ in the matter era, which 
corresponds to the tracker solution for 
the scalar Galileon \cite{DeFelice:2010pv,DeFelice:2010nf}. 
Existence of the negative coupling $Q$ allows 
an interesting possibility for reducing such tensions. 
As we will show in this paper, 
the no-ghost condition of the matter sector 
on future de Sitter solutions places the upper limit 
on $|Q f(X)|$ as well as the maximum value 
of $w_{\rm DE}$ during the matter dominance.  
In particular, the model with $p_3=2$ can be compatible with observational 
data of the background expansion history.

We note that the negative coupling $Q$ corresponds to 
the decay of dark matter to dark energy. 
There are other phenomenological coupled dark energy models 
in which the decay of dark matter can reduce the tensions of 
today's Hubble constant $H_0$ and the amplitude of matter 
perturbations $\sigma_8$ between CMB and low-redshift 
measurements \cite{DiValentino:2017iww,DiValentino:2019ffd}. 
This gives us the further motivation to compute the effective gravitational 
coupling for CDM in coupled vector dark energy and to confront the model with observations. 
We will address these issues in a future separate 
publication.

This paper is organized as follows.
In Sec.~\ref{eomsec}, we derive the background equations 
of motion and discuss how the vector field and 
CDM are coupled to each other.
In Sec.~\ref{stasec}, we identify conditions for the absence 
of ghosts and Laplacian instabilities of tensor, vector, 
and scalar perturbations for linear cosmological perturbations.
In Sec.~\ref{modelsec}, we propose a concrete coupled vector 
model of late-time cosmic acceleration and study the dynamics of 
dark energy together with the stability conditions. 
Sec.~\ref{concludesec} is devoted to conclusions.

Throughout the paper, the Greek and Latin indices are used to 
represent four-and three-dimensional quantities, respectively. 
For the partial and covariant derivatives with respect to $x^{\mu}$, 
we adopt the notations $\partial_{\mu}$ and $\nabla_{\mu}$, respectively. 
We also use the natural unit in which the 
speed of light $c$ and the reduced Planck constant $\hbar$ 
are equivalent to 1.
The capital label ``$I$'' represents 
different matter species (CDM, baryons, radiations). 

\section{Coupled vector dark energy model 
and background equations}
\label{eomsec}

We study the cosmology of cubic-order GP theories \cite{Heisenberg,Tasinato1,Tasinato2}
in which a vector field $A_{\mu}$ is coupled to CDM. 
We also take baryons 
and radiations into account and assume that they are not
directly coupled to $A_{\mu}$.
The total action is then given by 
\be
{\cal S} = {\cal S}_{\rm GP}
+{\cal S}_M
+{\cal S}_{\rm int}\,,
\label{GPaction}
\ee
where
\be
{\cal S}_{\rm GP}=
\int {\rm d}^{4}x \sqrt{-g} 
\left[ \frac{M_{\rm pl}^{2}}{2} R +G_{2}(X,\, F) 
+ G_{3}(X) \nabla_{\mu} A^{\mu} \right]\,,
\label{SGP}
\ee
with $g$ being the determinant of metric tensor $g_{\mu\nu}$.
The function $G_2$ depends on $X= -A_{\mu} A^{\mu}/2$ and 
$F=-F_{\mu \nu} F^{\mu \nu}/4$, while $G_3$ is a function 
of $X$ alone. 
The massive vector field with the standard Maxwell Lagrangian 
corresponds to $G_2(X, F)=m^2 X+F$, where the mass squared 
$m^2$ can be either positive or negative. 
Existence of the Lagrangian $G_{3}(X) \nabla_{\mu} A^{\mu}$ allows 
the possibility for realizing a de Sitter solution with 
constant $X$ \cite{DeFelice:2016yws,DeFelice:2016uil}. 
Besides two tensor polarizations arising from the Ricci scalar $R$, 
there are two transverse vector modes and one longitudinal scalar 
arising from the breaking of $U(1)$ gauge symmetry.

For the matter action ${\cal S}_M$, we consider perfect fluids 
described by the Schutz-Sorkin action \cite{Sorkin,Brown,DGS}
\be
{\cal S}_{M} = - \sum_{I=c,b,r}\int {\rm d}^{4}x \Bigl[
\sqrt{-g}\,\rho_I(n_I)
+ J_I^{\mu} \bigl( \partial_{\mu} \ell_I 
+ \mathcal{A}_{I1} \partial_{\mu} \mathcal{B}_{I1} 
+ \mathcal{A}_{I2} \partial_{\mu} \mathcal{B}_{I2} \bigr)
\Bigr]\,,
\label{Schutz}
\ee
where the subscripts $I=c,b,r$ represent CDM, baryons, 
and radiations, respectively.
The energy density $\rho_I$ depends on the  
fluid number density $n_I$. 
We note that the perturbation $\delta \rho_I$  of energy 
density plays the role of a dynamical scalar degree of freedom 
in the matter sector.
The vector field $J_I^{\mu}$ is related to $n_I$ 
according to 
\be
n_I=\sqrt{\frac{J_I^{\mu} J_I^{\nu}
g_{\mu \nu}}{g}}\,.
\label{ndef}
\ee
The vector field $J_I^{\mu}$ is related to the four-velocity 
of each matter, whereas the  scalar field $\ell_I$ is a Lagrange 
multiplier corresponding to a constraint of the particle conservation.
The quantities ${\cal A}_{1,2}$ and   ${\cal B}_{1,2}$  are the Lagrange multipliers 
and Lagrange coordinates of fluids, respectively, both of which are associated with 
nondynamical intrinsic vector modes.
In Sec.~\ref{stasec}, we vary the second-order actions of vector and scalar perturbations 
with respect to these nondynamical variables 
and eliminate them from 
the corresponding actions. This is for the purpose of deriving stability 
conditions of dynamical vector and scalar degrees of freedom.

The energy-momentum tensor of each perfect fluid 
is given by 
\be
(T_I)^{\mu}_{\nu}=\left( \rho_I+P_I \right) 
u_I^{\mu} u_{I\nu}+P_I \delta^{\mu}_{\nu}\,,
\label{Tmunu}
\ee
where $\rho_I$ and $P_I$ correspond to 
the energy density and pressure, respectively, 
and $u_{I{\mu}}$ is the four-velocity related to 
$J_{I{\mu}}$, as 
\be
u_{I{\mu}} \equiv \frac{J_{I{\mu}}}{n_I\sqrt{-g}}\,.
\label{udef}
\ee
{}From Eq.~(\ref{ndef}), it follows that 
\be
u_I^{\mu} u_{I{\mu}}=-1\,.
\label{ure}
\ee

We consider the case in which both baryons and radiations 
are uncoupled to the vector field. 
Even in this case, the cubic interaction 
$G_3(X) \nabla_{\mu}A^{\mu}$ leads to the gravitational coupling
for baryons different from the Newton gravitational constant for linear 
cosmological perturbations \cite{Nakamura:2018oyy}. In over-density regions of the Universe, 
however, the fifth force is suppressed by the same cubic interaction  
through the operation of the Vainshtein mechanism \cite{DeFelice:2016cri}.

On the other hand, we assume that $A_{\mu}$ is coupled to  
CDM with the interacting action
\be
{\cal S}_{\rm int}= \int {\rm d}^4x \sqrt{-g}\,
Q\,f(X)\,T_{c}\,,
\label{Acoupling}
\ee
where $Q$ is a dimensionless coupling constant, 
$f$ is a function of $X$, and $T_{c}$ is the trace $(T_c)^{\mu}_{\mu}$ 
of Eq.~(\ref{Tmunu}). 
We focus on the case in which the CDM pressure vanishes, i.e., 
\be
P_c=0\,.
\label{Pc}
\ee
On using Eq.~(\ref{Tmunu}) with Eq.~(\ref{ure}), 
the action (\ref{Acoupling}) reduces to
\be
{\cal S}_{\rm int}=-\int {\rm d}^4x \sqrt{-g}\,
Q\,f(X)\,\rho_c (n_c)\,.
\label{Acoupling2}
\ee

Taking the variation of the action (\ref{GPaction}) with 
respect to $J_I^{\mu}$ and employing the relation 
$\partial n_I/\partial J_I^{\mu}
=J_{I{\mu}}/(n_I g)$, we obtain 
\be
\partial_{\mu} \ell_{I}+\mathcal{A}_{I1} \partial_{\mu} 
\mathcal{B}_{I1} 
+ \mathcal{A}_{I2} \partial_{\mu} \mathcal{B}_{I2}
= u_{I{\mu}} \rho_{I,n_I}\,, \qquad \quad {\rm for}~I=b,r\,,
\label{l1}
\ee
where $\rho_{I,n_I} \equiv \partial \rho_I/\partial n_I$.
For CDM, we have
\be
\partial_{\mu} \ell_{c}+\mathcal{A}_{c1} \partial_{\mu} 
\mathcal{B}_{c1} 
+ \mathcal{A}_{c2} \partial_{\mu} \mathcal{B}_{c2}
= u_{c{\mu}} \rho_{c,n_c} \left[1+Qf(X) \right] \,.
\label{l2}
\ee

Let us consider the flat FLRW background described by the line element 
\be
{\rm d} s^{2} = - N^2(t) {\rm d}t^2
+a^2(t) \delta_{ij} {\rm d}x^{i} {\rm d}x^{j}\,,
\label{metric}
\ee
where $N(t)$ is the lapse, $a(t)$ is the scale factor, 
and $t$ is the cosmic time.
The vector field profile compatible with the line element
(\ref{metric}) is given by 
\be
A^{\mu}=\left( \frac{\phi(t)}{N(t)},0,0,0 
\right)\,,
\ee
where $\phi$ depends on $t$. 
Then, we have that $X=\phi^2(t)/2$ and $F=0$.
The temporal vector component $\phi(t)$ is an auxiliary field 
playing the role of dark energy at the background 
level \cite{DeFelice:2016yws,DeFelice:2016uil}. 
Replacing $A_{\mu}$ for $\partial_{\mu}\varphi$, the  
GP theories discussed below have the analogue to shift-symmetric 
scalar-tensor theories with the scalar field $\varphi$. 
At the background level, integrating the relation 
$A_{0}=\partial_{0}\varphi$ gives the additional integration 
constant for $\varphi$. In other words, the scalar-tensor counterpart 
of GP theories has an additional degree of freedom for the choice 
of initial conditions \cite{DeFelice:2010pv,DeFelice:2010nf}. This means that the background dynamics 
in GP theories is not generally the same as the corresponding 
analogue of scalar-tensor theories.

Up to boundary terms, the action (\ref{SGP}) of the gravity 
and vector-field sectors yields
\be
{\cal S}_{\rm GP}=\int {\rm d}^4 x\,a^3 \left( N G_2+G_3 \dot{\phi}
+3G_3H \phi -\frac{3M_{\rm pl}^2 H^2}{N} \right)\,,
\label{Scon1}
\ee
where a dot represents the derivative with respect to $t$, 
and $H \equiv \dot{a}/a$ is the Hubble expansion rate.

Since the fluid four-velocity in its rest frame is given by 
$u_I^{\mu}=(N^{-1},0,0,0)$, Eq.~(\ref{udef}) leads to
\be
J_I^0=n_I a^3\,.
\label{nJ}
\ee
Varying the action (\ref{Schutz}) with respect to $\ell_I$, 
we obtain 
\be
{\cal N}_I \equiv J_I^0=n_I a^3={\rm constant}\,,
\label{Ji}
\ee
which means that the particle number ${\cal N}_I$ is conserved. 
In other words, the number density of each matter 
species (including CDM) obeys 
\be
\dot{n}_I+3H n_I=0\,.
\label{ni}
\ee

On using Eq.~(\ref{nJ}), the action in the matter sector 
reduces to 
\be
{\cal S}_M+{\cal S}_{\rm int}
=-\int {\rm d}^4 x \left[ N a^3 \left\{ [1+Qf(X)]\rho_c
+\rho_b+\rho_r \right\}+a^3 \left( n_c \dot{\ell}_c 
+ n_b \dot{\ell}_b+ n_r \dot{\ell}_r \right) \right]\,. 
\label{Scon2}
\ee
On the FLRW background (\ref{metric})
the vector modes are absent in Eqs.~(\ref{l1})-(\ref{l2}), 
with the four velocity $u_{I{\mu}}=(-N,0,0,0)$,
Then, we obtain
\ba
\dot{\ell}_I &=& -N\rho_{I,n_I}\,,\qquad \quad 
{\rm for}~I=b,r\,,\label{lre1}\\
\dot{\ell}_c &=& -N\rho_{c,n_c} \left[1+Qf(X) 
\right]\,.\label{lre2}
\ea
The pressure $P_I$ associated with the energy density 
$\rho_I$ is given by \cite{DeFelice:2016yws,DeFelice:2016uil}
\be
P_I \equiv n_I \rho_{I,n_I}-\rho_I\,.
\label{Pi}
\ee
Substituting Eqs.~(\ref{lre1}) and (\ref{lre2}) into 
Eq.~(\ref{Scon2}) and taking the limit $Q\to 0$, 
the action (\ref{Scon2}) reduces to 
the sum of pressures, i.e., 
$\int {\rm d}^4 x\sqrt{-g}\sum_{I=c,b,r}P_I$. 
For $Q \neq 0$, the additional term $Qf(X)$ is 
present for CDM.

On using Eqs.~(\ref{ni}), (\ref{Pi}) and the property 
$\dot{\rho}_{I}(n_I) = \rho_{I,n_I} \dot{n}_I$, 
the energy density $\rho_I(n_I)$ of 
each matter component obeys
\be
\dot{\rho}_I+3H \left( \rho_I+P_I 
\right)=0\,.
\label{coneq}
\ee
We consider the case in which the pressures of baryons 
and radiations satisfy $P_b=0$ and $P_r=\rho_r/3$, 
respectively, with the vanishing CDM pressure (\ref{Pc}).
Then, the energy densities of three matter 
components obey
\ba
& &
\dot{\rho}_b+3H \rho_b=0\,,
\label{rhobeq}\\
& &
\dot{\rho}_r+4H \rho_r=0\,,
\label{rhoreq}\\
& &
\dot{\rho}_c+3H \rho_c=0\,.
\label{rhoceq}
\ea

Varying the sum of actions (\ref{Scon1}) and (\ref{Scon2}) with 
respect to $N$, $a$, $\phi$ and setting $N=1$ at the end, 
we obtain the background equations:
\ba
& & 3 M_{\rm pl}^{2} H^{2} = - G_{2} 
+\left( 1+Q f \right) \rho_{c} 
+\rho_{b}+\rho_{r} \,,
\label{Eq00}\\
& & M_{\rm pl}^{2} \left( 2 \dot{H}+ 3H^{2} 
\right) = -G_{2} + G_{3,X} \phi^{2} \dot{\phi} 
-\frac{1}{3}\rho_r \,, 
\label{Eq11}\\
& &\phi \left( G_{2,X} + 3 G_{3,X} H \phi -Qf_{,X}\rho_{c} 
\right) = 0 \,,
\label{EqA0}
\ea
where $G_{i,X} \equiv \partial G_i/\partial X$.
We focus on the branch $\phi \neq 0$ in Eq.~(\ref{EqA0}), i.e.,
\be
G_{2,X} + 3 G_{3,X} H \phi -Qf_{,X}\rho_{c}=0\,.
\label{EqA0d}
\ee
We define the energy density of CDM containing the effect of 
interactions with the vector field, such that 
\be
\tilde{\rho}_c \equiv 
\left( 1+Q f \right) \rho_{c}\,.
\ee
On using Eq.~(\ref{rhoceq}), we find that $\tilde{\rho}_c$ 
obeys the differential equation:
\be
\dot{\tilde{\rho}}_c+3H \tilde{\rho}_c
=\frac{Q f_{,X}\phi \dot{\phi}}{1+Qf} 
\tilde{\rho}_c\,.
\label{rhoccon}
\ee
Unlike the conserved CDM density $\rho_c$, the effective CDM density 
$\tilde{\rho}_c$ is a physical quantity whose continuity equation contains
the effect of couplings on the right hand side of Eq.~(\ref{rhoccon}). 
In spite of the conservation of CDM particle number (\ref{ni}), i.e., 
$n_c a^3={\rm constant}$, the CDM density $\tilde{\rho}_c$ acquires 
the effective mass term $m_c(Q_f)$ by the coupling $Q_f=Qf$, such that 
$\tilde{\rho}_c=m_c(Q_f) n_c$. 
This means that, unlike $\rho_c$, the effective density $\tilde{\rho}_c$ does not 
obey the standard continuity equation. We also observe that the matter action 
(\ref{Scon2}) contains the term proportional to $\tilde{\rho}_c$.

We also define the energy density $\rho_{\rm DE}$ 
and pressure $P_{\rm DE}$ of 
dark energy (arising from the vector field), as
\ba
\rho_{\rm DE} &=& -G_2\,,\label{rhode}\\
P_{\rm DE} &=& G_2-G_{3,X} \phi^{2} \dot{\phi}\,,
\label{Pde}
\ea
with the equation of state
\be
w_{\rm DE} \equiv \frac{P_{\rm DE}}{\rho_{\rm DE}}
=-1+\frac{G_{3,X} \phi^{2} \dot{\phi}}{G_2}\,.
\label{wDEdef}
\ee
Taking the time derivative of Eq.~(\ref{rhode}) 
and using Eqs.~(\ref{EqA0d}) and (\ref{Pde}), 
it follows that 
\be
\dot{\rho}_{\rm DE}+3H \left( \rho_{\rm DE}
+P_{\rm DE} \right)=
-\frac{Q f_{,X}\phi \dot{\phi}}{1+Qf} 
\tilde{\rho}_c\,.
\label{rhodecon}
\ee
Comparing Eq.~(\ref{rhoccon}) with Eq.~(\ref{rhodecon}), it is clear 
that the vector field and CDM interact with each other 
through the couplings with opposite signs.

\section{Conditions for avoiding ghosts and gradient instabilities}
\label{stasec}

We derive conditions for the absence of ghosts and Laplacian instabilities 
for tensor, vector, and scalar perturbations. 
Throughout the paper, we focus on the evolution of linear cosmological perturbations 
without considering the nonlinear regime, e.g., the small-scale region in which dark matter 
is concentrated in 
halos of galaxies. Even for nonlinear perturbations today, they were in the linear 
regime in the past cosmic growth history. The stability conditions derived 
in this section  should be consistently satisfied to ensure 
the stability of perturbations from the past to today
for linear perturbations deep inside the sound horizons.

We consider the perturbed line element
in the flat gauge on the flat FLRW background:
\be
{\rm d}s^{2} = - \left( 1 + 2 \alpha \right) {\rm d}t^{2}
+ 2 \left( \partial_{i}\chi + V_{i} \right) {\rm d}t\,{\rm d}x^{i}
+ a^{2} \left( \delta_{ij} + h_{ij} \right) {\rm d}x^{i} {\rm d}x^{j}\,,
\label{permet}
\ee
where $\alpha$ and $\chi$ are scalar perturbations, $V_i$ is the vector perturbation 
satisfying the transverse condition $\partial^i V_i=0$, and $h_{ij}$ is the tensor 
perturbation obeying the transverse and traceless conditions $\partial^i h_{ij}=0$ and $h^i_i=0$. 
All the perturbed quantities depend on both $t$ and $x^i$. 

We decompose the temporal and spatial components of 
$A^{\mu}=(A^0, A^i)$ into the background and perturbed parts:
\be
A^0=\phi(t)+\delta \phi\,,\qquad 
A^i=\frac{1}{a^{2}(t)} \delta^{ij} 
\left( \partial_{j} \chi_{V} + E_{j}\right)\,,
\ee
where $\delta\phi$ and $\chi_{V}$ are scalar perturbations, and 
$E_{j}$ is the vector perturbation obeying the transverse condition
$\partial^{j}E_{j} = 0$.
Similarly, the temporal and spatial components of 
the vector $J_I^{\mu}=(J_I^{0},J_I^{i})$ 
(with $I=b,r,c$) in the Schutz-Sorkin action (\ref{Schutz})
are decomposed as
\be
J_I^{0}={\cal N}_I(t)+\delta J_I\,,\qquad
J_I^{i}= \frac{1}{a^2(t)} \delta^{ij} 
\left( \partial_j \delta j_I+W_{Ij} \right)\,,
\ee
where $\delta J_I$ and $\delta j_I$ are scalar perturbations, 
and $W_{Ij}$ is the vector perturbation obeying the 
transverse condition $\partial^{j}W_{Ij}=0$.
We express the quantities $\ell_{b,r,c}$ in the form
\ba
\ell_I
&=&-\int^t \rho_{I,{n_I}}(\tilde{t})\,{\rm d}\tilde{t}
-\rho_{I,{n_I}}(t)\,v_I\,,\qquad 
{\rm for}~I=b,r\,,\label{ell0}\\
\ell_c
&=&-\int^t \left[ 1+Qf(\tilde{t}) \right]
\rho_{c,{n_c}}(\tilde{t})\,{\rm d}\tilde{t}
-\left[ 1+Qf(t) \right] 
\rho_{c,{n_c}}(t)\,v_c\,,
\label{ell}
\ea
where $\rho_{I,{n_I}}(t)$ and $Qf(t)$ are evaluated on the background, 
$v_I$ and $v_c$ are velocity potentials having 
the dependence of both $t$ and $x^i$. 
We observe that the background Eqs.~(\ref{lre1}) and (\ref{lre2})
with $N=1$ are consistent with
Eqs.~(\ref{ell0}) and (\ref{ell}).

As for the vector perturbations $V_{i}$, $E_{i}$ and $W_{Ii}$, 
we choose 
\be
V_{i} = \left( V_{1}(t,z),\,V_{2}(t,z),\,0 \right) \,,\qquad
E_{i} = \left( E_{1}(t,z),\,E_{2}(t,z),\,0 \right) \,,\qquad
W_{I i} = \left( W_{I1}(t,z),\,W_{I2}(t,z),\,0 \right) \,,
\ee
whose $x$ and $y$ components depend on $t$ and $z$.
They are consistent with the transverse conditions 
mentioned above.
For the quantities ${\cal A}_{Ii}$ and ${\cal B}_{Ii}$ 
in Eq.~(\ref{Schutz}), we choose
\ba
& &
{\cal A}_{I1} = \delta{\cal A}_{I1}(t,z)\,, \quad
{\cal A}_{I2} = \delta{\cal A}_{I2}(t,z)\,, \label{dAdef}\\
& &{\cal B}_{I1} = x + \delta{\cal B}_{I1}(t,z)\,, \quad
{\cal B}_{I2} = y + \delta{\cal B}_{I2}(t,z)\,, \label{dBdef}
\ea
where $\delta{\cal A}_{Ii}$ and $\delta{\cal B}_{Ii}$ 
are linearly perturbed quantities. 
We recall that the four-velocities of baryons and radiations 
have the relation (\ref{l1}), while the CDM four-velocity 
satisfies the relation (\ref{l2}).
On defining the intrinsic velocity vectors $v_{Ii}$ as
\ba
& &
\delta{\cal A}_{Ii} = \rho_{I,n_I}(t)\,v_{Ii}\,,
\qquad {\rm for}~I=b,r\,,\label{dAI} \\
& &
\delta{\cal A}_{ci} = \left[ 1 + Q f(t) 
\right] \rho_{c,n_c}(t)\,v_{ci}\,,
\label{dAc} 
\ea
the spatial components of $u_{I{\mu}}$ yield
\be
u_{Ii} = -\partial_{i}v_{I} + v_{Ii}\,,
\qquad {\rm for}~I=b,r,c\,,
\label{u_i}
\ee
where $v_{b,r,c}$ are the scalar velocity 
potentials appearing in Eqs.~(\ref{ell0}) and (\ref{ell}). 

\subsection{Tensor perturbations}

The tensor perturbation is expressed in terms of 
the sum of two polarization modes, as 
$h_{ij} = h_{+}e_{ij}^{+} + h_{\times} e_{ij}^{\times}$. 
In Fourier space with the comoving wavenumber $\bm{k}$, 
the unit vectors $e_{ij}^{+}$ and $e_{ij}^{\times}$ 
satisfy the normalizations 
$e_{ij}^{+}(\bm{k})e_{ij}^{+}(-\bm{k})^{*}=1$, 
$e_{ij}^{\times}(\bm{k})e_{ij}^{\times}(-\bm{k})^{*}=1$, and 
$e_{ij}^{+}(\bm{k})e_{ij}^{\times}(-\bm{k})^{*}=0$.

Expanding the action (\ref{GPaction}) up to second order in $h_{ij}$ 
and using the background Eq.~(\ref{Eq11}), it follows that 
the terms containing $h_{+}^{2}$ and $h_{\times}^{2}$ 
identically vanish. The resulting second-order action 
of tensor perturbations is given by 
\be
{\cal S}_{T}^{(2)} = \sum_{\lambda=+,\times}
\int {\rm d}^4 x\,
a^{3} \frac{q_{T}}{8} 
\left[\dot{h}_{\lambda}^{2} - \frac{c_{T}^{2}}{a^{2}} (\partial h_{\lambda})^{2}\right]\,,
\label{ST}
\ee
where
\be
q_{T} = M_{\rm pl}^{2} \,, \qquad
c_{T}^{2} = 1\,.
\ee
The action (\ref{ST}) is the same as that in 
general relativity. 
Hence the propagation of tensor perturbations 
is not modified by the nonvanishing coupling $Q$.
Since the speed $c_T$ of gravitational waves is equivalent to 1, 
the coupled dark energy theories given by (\ref{GPaction}) are 
consistent with the bound 
arising from the GW170817 event \cite{GW170817}.

\subsection{Vector perturbations}

As for vector perturbations, we first expand the actions 
(\ref{Schutz}) and (\ref{Acoupling}) up to second order. 
The resulting quadratic-order actions are given, respectively, by  
\begin{align}
({\cal S}_{M}^{(2)})_{V} &= 
\int {\rm d}^4 x\sum_{I=b,r,c}\sum_{i=1}^{2} 
\left[ \frac{1}{2a^{2}} \Bigl\{ 
\frac{\rho_{I,n_I}}{{\cal N}_{I}} 
\left( W_{Ii}^{2} + {\cal N}_{I}^{2} V_{i}^{2}\right)
+  2 \rho_{I,n_I} V_{i} W_{Ii} - a^{3} \rho_{I} V_{i}^2 
\Bigr\}
- {\cal N}_{I} \delta{\cal A}_{Ii} \dot{\delta{\cal B}_{Ii}}
- \frac{1}{a^{2}} W_{Ii} \delta{\cal A}_{Ii}
\right]\,,\label{SMva1} \\
({\cal S}_{\rm int}^{(2)})_{V} &= 
\int {\rm d}^4 x \sum_{i=1}^{2} 
Q\left[ \frac{f}{2 a^{2}} \left\{
\frac{\rho_{c,n_c}}{{\cal N}_{c}} (W_{ci}^{2}
+{\cal N}_{c}^{2} V_{i}^{2})
+ 2  \rho_{c,n_c} V_{i} W_{ci}
- a^{3}  \rho_{c} V_{i}^{2} \right\}
+\frac{a f_{,X} \rho_{c}}{2}
(E_{i}+2\phi V_{i}) E_{i}
\right]\,.\label{SMva2}
\end{align}
Varying the action  
$({\cal S}_{M}^{(2)})_{V}+({\cal S}_{\rm int}^{(2)})_{V}$ 
with respect to $W_{Ii}$ and using Eqs.~(\ref{dAI}) and 
(\ref{dAc}), it follows that
\be
W_{Ii}={\cal N}_i \left( v_{Ii}-V_i \right)\,, 
\ee
which hold for $I=b,r,c$. 
Substituting this relation into Eqs.~(\ref{SMva1})-(\ref{SMva2})
and varying $({\cal S}_{M}^{(2)})_{V}+({\cal S}_{\rm int}^{(2)})_{V}$ 
with respect to $v_{Ii}$ and $\delta {\cal B}_{Ii}$, we obtain
\ba
v_{Ii} &=&
V_i-a^2 \dot{\delta {\cal B}}_{Ii}\,,\\
\delta{\cal A}_{Ii} &=& C_{Ii}\,,
\ea
where $C_{Ii}$ are constants in time.

After integrating out the perturbations $W_{Ii}$ and $\delta{\cal A}_{Ii}$, 
the resulting second-order action in the matter sector yields
\be
({\cal S}_{M}^{(2)})_{V} + ({\cal S}_{\rm int}^{(2)})_{V} 
=\int {\rm d}^4 x\,\frac{a}{2} \sum_{i=1}^2 \left[ \sum_{I=b,r,c} 
\left( n_I \rho_{I,n_I} v_{Ii}^2-\rho_I V_i^2 \right)
+Q \left\{ \left( n_c \rho_{c,n_c} v_{ci}^2-\rho_c V_i^2 
\right)f+\rho_c f_{,X} E_i (2\phi V_i+E_i) \right\}
\right].
\ee
We now expand the action (\ref{GPaction}) up to second order 
in vector perturbations. 
In doing so, it is convenient to introduce the combination
\be
Z_{i} \equiv E_{i} + \phi(t) V_{i}\,,
\ee
which correspond to $A_i$.
Then, the total quadratic-order action 
for vector perturbations yields
\be
{\cal S}_{V}^{(2)} = \int {\rm d}^4x \sum_{i=1}^{2} \frac{a}{2} \left[
q_{V} \dot{Z}_{i}^{2} - \frac{q_{V}}{a^{2}} (\partial Z_{i})^{2}
- G_{3,X} \dot{\phi} Z_{i}^{2} 
+ \frac{q_{T}}{2 a^{2}} (\partial V_{i})^{2}
+ \rho_{b} v_{bi}^{2}
+ \frac{4}{3} \rho_{r} v_{ri}^{2}
+ \left( 1 + Q f \right) \rho_{c} v_{ci}^{2}
\right]\,,
\label{SV2_v2}
\ee
where
\be
q_{V} \equiv G_{2,F}\,.
\ee
Varying the action (\ref{SV2_v2}) with respect to $V_i$ 
in Fourier space with the comoving 
wavenumber $k=|{\bm k}|$, we obtain
\be
\frac{q_{T}}{2} a k^2 V_{i} =
-\left( {\cal N}_b C_{bi}
+{\cal N}_r C_{ri}+{\cal N}_c C_{ci} \right)\,,
\ee
which can be used to eliminate the fourth term in Eq.~(\ref{SV2_v2}).
For linear perturbations deep inside the Hubble radius, 
the action (\ref{SV2_v2}) reduces to
\be
{\cal S}_{V}^{(2)} \simeq \sum_{i = 1}^{2}\int {\rm d}^4 x\,
\frac{a}{2}q_V \left( \dot{Z}_{i}^{2}-c_V^2 \frac{k^{2}}{a^{2}} Z_{i}^{2} \right)\,,
\label{SV2}
\ee
where 
\be
c_V^2=1\,.
\ee
The two dynamical fields $Z_{1}$ and $Z_{2}$
propagate with the  speed $c_V$ equivalent to 1, 
so there are no Laplacian instabilities of vector perturbations.
The no-ghost condition corresponds to $q_V>0$, i.e.,
\be
G_{2,F}>0\,.
\label{veccon}
\ee
{}From the above discussion, it is clear that the coupling $Q$ 
does not affect the stability conditions of linear vector perturbations.

\subsection{Scalar perturbations}
\label{scalarsec}

To study the propagation of scalar perturbations, we first define 
the density perturbation $\delta \rho_I$ of each matter 
fluid ($I=c,b,r$), as
\be
\delta \rho_I \equiv \frac{\rho_{I,n_I}}{a^3} \delta J_I\,,
\ee
where $\rho_{I,n_I}$ solely depends on the number density 
$n_I$.
By defining $\delta \rho_I$ in this way, the perturbation of number 
density $n_I$, expanded up to second order, can be expressed as
\be
\delta n_I=\frac{\delta \rho_I}{\rho_{I,n_I}}
-\frac{({\cal N}_I \partial \chi+\partial \delta j_I)^2}
{2{\cal N}_I a^5}\,,
\ee
whose first term on the right hand side is 
consistent with the left hand side. 

Expanding ${\cal S}_{M}+{\cal S}_{\rm int}$ up 
to second order in scalar perturbations and varying the resulting 
quadratic-order action with respect to $\delta j_I$, 
it follows that 
\ba
\partial \delta j_{I} 
&=& -\mathcal{N}_{I} \left( \partial \chi + \partial v_{I} \right) \,,
\qquad {\rm for}~~I=c,b,r\,,
\label{deljre}
\ea
which can be used to eliminate the nondynamical fields
$\delta j_{I}$ from the matter action. 
We note that the relations (\ref{deljre}) also follow from 
the spatial components of Eqs.~(\ref{l1}) and (\ref{l2}).

The propagation speed squares of matter perfect fluids 
are defined as
\be
c_I^2=\frac{n_I \rho_{I,n_I n_I}}{\rho_{I,n_I}}\,,
\ee
with $I=c,b,r$.
We focus on the case in which $c_I^2$ for CDM, 
baryons, and radiations are given, respectively, by 
\be
c_c^2=0\,,\qquad 
c_b^2=0\,,\qquad c_r^2=\frac{1}{3}\,.
\ee

To expand the action ${\cal S}_{\rm GP}$ up to quadratic order 
in scalar perturbations, we introduce the combination
\be
\psi \equiv \chi_V+\phi(t) \chi\,,
\ee
so that $A_i=\partial_i\psi$ in the scalar sector.
Using the background Eqs.~(\ref{Eq00})-(\ref{EqA0}), 
the second-order action arising from (\ref{GPaction}) reads
\be
{\cal S}_S^{(2)}={\cal S}_{Q=0}^{(2)}
+{\cal S}_Q^{(2)}\,,
\label{S2}
\ee
where 
\ba
{\cal S}_{Q=0}^{(2)} &=&\int {\rm d}^4 x
\,a^{3}\,\Biggl\{\sum_{I=c,b,r} \left\{
\left[ n_I\rho_{I,n_I}\,\frac{\partial^{2}\chi}{a^2}-\dot{\delta\rho}_I
-3H\left(1+c_I^2 \right)\,\delta\rho_I \right]v_I
-\frac{n_I \rho_{I,n_I}}{2}\,\frac{(\partial v_I)^{2}}{a^{2}}
-\frac{c_I^2}{2n_I \rho_{I,n_I}}(\delta \rho_I)^{2}
-\alpha \delta\rho_I \right\}
\nonumber \\
 &  & {}-w_{3}\,\frac{(\partial\alpha)^{2}}{a^{2}}+w_{4}\alpha^{2}
 -\left[(3Hw_{1}-2w_{4})\frac{\delta\phi}{\phi}-w_{3}\,\frac{\partial^{2}(\delta\phi)}
 {a^{2}\phi}-w_{3}\,\frac{\partial^{2}\dot{\psi}}{a^{2}\phi}
 +w_{6}\,\frac{\partial^{2}\psi}{a^{2}}\right] \alpha 
 -\frac{w_{3}}{4}\,\frac{(\partial\delta\phi)^{2}}{a^{2}\phi^{2}}
 +w_{5}\,\frac{(\delta\phi)^{2}}{\phi^{2}}\nonumber \\
 &  & 
 -\left[\frac{(w_{6}\phi+w_{2})\psi}{2}-\frac{w_{3}}{2}\dot{\psi}\right]
 \frac{\partial^{2}(\delta\phi)}{a^{2}\phi^{2}}
 -\frac{w_{3}}{4\phi^{2}}\,\frac{(\partial\dot{\psi})^{2}}{a^{2}}+\frac{w_{7}}{2}\,
 \frac{(\partial\psi)^{2}}{a^{2}}
 +\left(w_{1}\alpha+\frac{w_{2}\delta\phi}{\phi}\right)\frac{\partial^{2}\chi}{a^{2}}\Biggr\}\,,
\ea
and 
\ba
{\cal S}_Q^{(2)}
&=&\int {\rm d}^4 x\,a^3 Q
\biggl[ 
\left( n_c\rho_{c,n_c}\,\frac{\partial^{2}\chi}{a^2}
-\dot{\delta\rho}_c
-3H\delta\rho_c \right) fv_c
-\frac{n_{c} \rho_{c,n_c}f}{2} 
\frac{(\partial v_{c})^{2}}{a^{2}}
- (f+f_{,X}\phi^2) \alpha \delta \rho_c \nonumber\\
& &\qquad \qquad \quad\,-f_{,X}\phi\,\delta \phi\,\delta \rho_c 
-\frac{1}{2}f_{,XX}\phi^2 \rho_c 
\left( \phi \alpha+\delta \phi \right)^2
\biggr]\,,
\ea
with 
\ba
w_{1} &=& - \phi^{3} G_{3,X} - 2 M_{\rm pl}^{2} H \,,\\
w_{2} &=& w_{1} + 2M_{\rm pl}^2 H=- \phi^{3} G_{3,X}  \,, \\
w_{3} &=& - 2 \phi^{2} q_{V} \,, \label{w3}\\
w_{4} &=& \frac{1}{2} \phi^{4} G_{2,XX}
- \frac{3}{2} H \phi^{3} (G_{3,X} - \phi^{2} G_{3,XX})
- 3 M_{\rm pl}^{2} H^{2} \,, \\
w_{5} &=& w_{4} - \frac{3}{2} H (w_{1} + w_{2}) \,, \\
w_{6} &=& \frac{1}{\phi} w_{2}=- \phi^{2} G_{3,X} \,, \\
w_{7} &=&\frac{\dot{\phi}}{\phi^3} w_{2}
=-\dot{\phi} G_{3,X} \,.
\ea
The action ${\cal S}_{Q=0}^{(2)}$ coincides with that derived in 
Refs.~\cite{DeFelice:2016yws,DeFelice:2016uil} in the single-fluid limit. 
The coupling $Q$ gives rise to the additional action ${\cal S}_{Q}^{(2)}$ 
to  ${\cal S}_{Q=0}^{(2)}$.
We note that the intrinsic vector mode affects the second-order scalar action 
through the quantity $w_3=-2\phi^2 q_V$ in Eq.~(\ref{w3}). 
Hence the scalar perturbation evolves differently compared to the corresponding 
analogue of scalar-tensor theories. Indeed, this difference manifests itself 
in the observations of ISW-galaxy cross-correlations \cite{Nakamura:2018oyy}.

Varying (\ref{S2}) with respect to nondynamical fields 
$\alpha$, $\chi$, $\delta\phi$, $v_b$, $v_r$, and $v_c$ 
in Fourier space, respectively, it follows that 
\ba
\hspace{-1cm}
& &
\sum_{I=c,b,r}\delta\rho_{I} 
- 2 w_{4} \alpha 
+ \left(3 H w_{1} - 2 w_{4} \right) \frac{\delta\phi}{\phi}
+ \frac{k^{2}}{a^{2}}\left( {\cal Y}+ w_{1} \chi- w_{6} \psi 
\right) = -Q( f + f_{,X}\phi^2) \delta \rho_{c}
-Q f_{,XX}\phi^3 \rho_c \left( \phi \alpha+\delta \phi \right),
\label{pereq1}\\
\hspace{-1cm}
& & \sum_{I=c,b,r} n_{I} \rho_{I,n_I} v_{I}
+ w_{1} \alpha + w_{2} \frac{\delta\phi}{\phi} 
=-n_c \rho_{c,n_c}Q f v_c\,,\\
\hspace{-1cm}
& &\left(3 H w_{1} - 2 w_{4}\right)\alpha -2 w_{5}\frac{\delta\phi}{\phi} 
+ \frac{k^{2}}{a^{2}} \left[
\frac{1}{2} {\cal Y}+w_2 \chi-\frac{1}{2} \left( 
\frac{w_2}{\phi}+w_6 \right)\psi
\right]= -Qf_{,X}\phi^2 \delta \rho_{c}
-Q f_{,XX}\phi^3 \rho_c \left( \phi \alpha+\delta \phi \right) \,,
\label{pereq3}\\
\hspace{-1cm}
& & \dot{\delta\rho_{I}}
+ 3 H \left( 1 + c_I^2 \right) \delta \rho_{I} 
+ \frac{k^{2}}{a^{2}} n_{I} \rho_{I,n_I} 
\left(\chi + v_I\right) = 0 \,,\qquad 
{\rm for}~~I=c,b,r\,,
\label{pereq4}
\ea
where 
\be
{\cal Y} \equiv \frac{w_3}{\phi} \left( 
\dot{\psi}+\delta \phi+2\phi \alpha \right)\,.
\ee

The dynamical perturbations correspond to the four fields 
$\psi$ and $\delta \rho_I$ ($I=c,b,r$). 
Under the gauge transformation $\tilde{t}=t+\xi^0$ and 
$\tilde{x}^{i}=x^{i}+\delta^{ij} \partial_j \xi$, 
these fields transform as $\tilde{\psi}=\psi+\phi\,\xi^{0}$ 
and $\widetilde{\delta \rho_I}=\delta \rho_I-\dot{\rho}_I \xi^0$, 
respectively.
If we consider two scalar metric perturbations $\zeta$ and $E$ 
in the spatial part of the line element (\ref{permet}), as 
in the form $a^2(t) [(1+2\zeta)\delta_{ij}+2\partial_i \partial_j E]{\rm d}x^i {\rm d}x^j$, 
they transform as $\tilde{\zeta}=\zeta-H \xi^0$ and 
$\tilde{E}=E-\xi$, respectively \cite{Heisenberg:2018wye}. 
The spatial gauge-transformation scalar $\xi$ 
is fixed by choosing $E=0$. 
{}From the temporal gauge-transformation, 
we can construct the gauge-invariant variables 
$\psi_{\zeta}=\psi+\phi \zeta/H$ and 
$\delta \rho_{I\zeta}=\delta \rho_I+3(\rho_I+P_I)\zeta$, 
where we used the continuity Eq.~(\ref{coneq}). 
Since $\zeta=0$ in the flat gauge, the 
perturbations $\psi_{\zeta}$ and $\delta \rho_{I\zeta}$ 
simply reduce to $\psi$ and $\delta \rho_I$, respectively.

Solving Eqs.~(\ref{pereq1})-(\ref{pereq4}) for $\alpha$, $\chi$, $\delta\phi$, $v_b$, $v_r$, $v_c$ 
and substituting them into Eq.~(\ref{S2}), 
the second-order action in Fourier space 
is expressed in the form
\be
{\cal S}_{S}^{(2)} = \int {\rm d}^4 x\, a^{3} \left(
 \dot{\vec{\mathcal{X}}}^{t}{\bm K}
\dot{\vec{\mathcal{X}}}
-\frac{k^2}{a^2}\vec{\mathcal{X}}^{t}{\bm G}
\vec{\mathcal{X}} 
 -\vec{\mathcal{X}}^{t}{\bm M}
\vec{\mathcal{X}}
-\vec{\mathcal{X}}^{t}{\bm B}
\dot{\vec{\mathcal{X}}}
\right) \,,
\label{SS}
\ee
where $\bm{K}$, $\bm{G}$, $\bm{M}$ and $\bm{B}$
are $4 \times 4$ matrices, 
and the vector field $\vec{\mathcal{X}}^{t}$ is given by
\be
\vec{\mathcal{X}}^{t} \equiv \left( \psi,\, \delta\rho_{c}/k,\, 
\delta\rho_{b}/k,\, \delta\rho_{r}/k \right)\,.
\ee
Neither $\bm{M}$ nor ${\bm B}$ contains the $k^2/a^2$ term. 
If there are the terms including the $k^2/a^2$ dependence in ${\bm B}$, 
it can be absorbed into $\bm{G}$ after the integration by parts.
For linear perturbations deep inside the sound horizon, the nonvanishing matrix 
components are
\ba
& &
K_{11}=\frac{H^{2} M_{\rm pl}^2 (3 w_{1}^{2} + 4 M_{\rm pl}^2 w_{4} - 2Q \rho_{c} M_{\rm pl}^2\,f_{,XX}\phi^4)}{\phi^{2} (w_{1} - 2 w_{2})^{2}} \,,\\
& &
K_{22}=\frac{a^{2} (1+Q f)}{2 n_{c} \rho_{c,n_c}} \,,\qquad
K_{33}=\frac{a^{2}}{2 n_{b} \rho_{b,n_b}} \,,\qquad
K_{44}=\frac{a^{2}}{2 n_{r} \rho_{r,n_r}} \,,
\ea
and 
\ba
& &G_{11}=\mathcal{G} + \dot{\mu} + H\mu
-\frac{w_2^2}{2(w_1-2w_2)^2 \phi^2} 
\left[ n_c \rho_{c,n_c} \left( 1+Qf \right)
+n_b \rho_{b,n_b}+n_r \rho_{r,n_r} \right]\,,\\
& &G_{22}=0 \,,\qquad
G_{33}=0 \,,\qquad
G_{44}=\frac{a^{2}c_r^2}{2 n_r\rho_{r,n_r}} \,,
\ea
where
\be
\mathcal{G} \equiv -\frac{4H^2M_{\rm pl}^4w_{2}^2}
{\phi^{2}w_{3}(w_{1}- 2 w_{2})^{2}} - \frac{\dot{\phi}}{2\phi^3}w_{2} \,,
\qquad \mu \equiv \frac{HM_{\rm pl}^2w_{2}}{\phi^{2}(w_{1}- 2 w_{2})} \,.
\ee
The scalar ghosts are absent for $K_{11}>0$, $K_{22}>0$, 
$K_{33}>0$, and $K_{44}>0$. 
Since $n_b \rho_{b,n_b}=\rho_b>0$ and $n_r \rho_{r,n_r}=4\rho_r/3>0$, 
the last two conditions trivially hold. 
The first condition is satisfied for 
\be
q_S \equiv 3 w_{1}^{2} + 4 M_{\rm pl}^2 w_{4} 
- 2Q \rho_{c} M_{\rm pl}^2 f_{,XX}\phi^4>0\,.
\label{qS}
\ee
Since  $n_c \rho_{c,n_c}=\rho_c>0$, the second condition 
translates to 
\be
q_c \equiv 1+Qf>0\,.
\label{qc}
\ee
For negative value of $Qf$, this gives the upper bound on $|Qf|$.

For linear perturbations deep inside the sound horizon, 
the second-order action (\ref{SS}) gives rise 
to the dispersion relation 
\be
{\rm det} \left( \omega^2 {\bm K}-\frac{k^2}{a^2} {\bm G} 
\right)=0\,,
\ee
with the frequency $\omega$. 
The scalar propagation speed $c_s$ is defined as 
$c_s^2=\omega^2 a^2/k^2$.
The propagation speed squared associated with the perturbation 
$\psi$ is given by $c_S^2=G_{11}/K_{11}$. To avoid the Laplacian 
instability, we require that 
\be
c_{S}^{2}=\frac{1}{K_{11}}
\left[\mathcal{G} + \dot{\mu} + H\mu
-\frac{w_2^2}{2(w_1-2w_2)^2 \phi^2} 
\left\{ \rho_c \left( 1+Qf \right)
+\rho_b+\frac{4}{3} \rho_r \right\} \right] \geq 0\,.
\label{cS}
\ee
The matter propagation speed squares, which correspond to 
the ratios $G_{22}/K_{22}$, $G_{33}/K_{33}$, $G_{44}/K_{44}$
for CDM, baryons, and radiations respectively, reduce to 
$c_c^2=0$, $c_b^2=0$, and $c_r^2=1/3$.
Hence there are no Laplacian instabilities 
for the three matter perfect fluids. 

The CDM perturbation $\delta \rho_c$ is associated 
with the perturbation of number-density dependent 
quantity $\rho_c(n_c)$, which is not identical to 
the perturbation $\tilde{\delta \rho_c}$ absorbing 
the contribution of coupling $Qf$ 
(related to the background CDM density $\tilde{\rho_c}=(1+Qf) \rho_c$).
In Eqs.~(\ref{pereq1}) and (\ref{pereq3}), 
however, we observe that $\delta \rho_c$ is coupled to the 
scalar perturbation $\psi$ arising from the vector field. 
Indeed, the effect of coupling appears 
in the expressions of $q_S$ and $c_S^2$ derived above. 
Unlike the background CDM density $\rho_c$ obeying Eq.~(\ref{rhoceq}), 
the interaction between $\delta \rho_c$ and $\psi$ manifests itself
at the level of linear perturbations.

\section{A concrete dark energy model}
\label{modelsec}

In this section, we propose a concrete coupled vector dark energy 
model and study the background cosmological 
dynamics by paying particular attention to the evolution 
of $w_{\rm DE}$. 
Let us consider the model given by the functions 
\be
G_2(X,F)=b_2 X+F\,,\qquad G_3(X)=b_3 X^{p_3}\,,\qquad 
f(X) = \left( \frac{X}{M_{\rm pl}^2} \right)^q\,,
\label{G23}
\ee
where $b_2$, $b_3$, $p_3$, and $q$ are constants.
Since $G_{2,F}=1$ in this model, the no-ghost condition 
(\ref{veccon}) of vector perturbations 
is automatically satisfied.

For the functions (\ref{G23}), Eq.~(\ref{EqA0d}) 
reduces to 
\be
b_2+3 \cdot 2^{1-p_3} b_3 p_3 H \phi^{2p_3-1}
-Qq \cdot 2^{1-q} M_{\rm pl}^{-2q} \rho_c 
\phi^{2(q-1)}=0\,.
\label{phieq}
\ee

For $Q=0$, there is the solution where $\phi$ solely 
depends on $H$, such that $\phi \propto H^{-1/(2p_3-1)}$. 
As long as the energy density of $A_{\mu}$ is 
subdominant to that of background fluids during the 
radiation and matter eras, the Hubble parameter evolves 
as $H \propto 1/t$ and hence the temporal vector component 
grows as $\phi \propto t^{1/(2p_3-1)}$ for $p_3>1/2$.
Finally, the solutions approach stable de Sitter attractors 
characterized by constant $\phi$ \cite{DeFelice:2016yws}.

For $Q \neq 0$, we would like to focus on the case 
where the third term in Eq.~(\ref{phieq}) is subdominant 
to the constant $b_2$ in the radiation era and 
temporally approaches a constant after the onset of matter dominance.
In doing so, we deal with the second term 
in Eq.~(\ref{phieq}) as a constant
during the matter era (as in the case $Q=0$), 
in which case $\phi \propto t^{1/(2p_3-1)}$. 
On using the property $\rho_c \propto a^{-3} \propto t^{-2}$ 
in this epoch, the third term in Eq.~(\ref{phieq}) is proportional
to $t^{(2q-4p_3)/(2p_3-1)}$.
For $q$ satisfying the relation 
\be
q=2p_3\,,
\label{q}
\ee
all the terms in Eq.~(\ref{phieq}) are constants 
during the matter era. 

For the choice (\ref{q}), the third term in Eq.~(\ref{phieq}) 
is subdominant to other two terms during the radiation dominance.
Indeed, exploiting the solutions $\phi \propto t^{1/(2p_3-1)}$ 
and $\rho_c \propto a^{-3} \propto t^{-3/2}$ in this epoch, 
the third term in Eq.~(\ref{phieq}) grows in proportion to 
$t^{1/2}$ toward a constant value in the matter era. 
After the onset of late-time cosmic acceleration, 
the coupling term in Eq.~(\ref{phieq}) starts to decrease 
toward 0 by reflecting the fact that $\rho_c$
decreases faster than $t^{-2}$. Finally, the 
solutions approach the de Sitter fixed point satisfying
\be
b_2+3 \cdot 2^{1-p_3} b_3 p_3 H_{\rm dS} 
\phi_{\rm dS}^{2p_3-1}=0\,,
\label{dS}
\ee
where the subscript ``dS'' represents the values 
on the de Sitter point. 

\subsection{Autonomous system}

In the following, we focus on the background cosmological 
dynamics for the power $q$ satisfying the relation (\ref{q}). 
In doing so, it is convenient to define 
\be
\Omega_{\rm DE} \equiv \frac{\rho_{\rm DE}}{3 M_{\rm pl}^{2} H^{2}}
= - \frac{b_{2} \phi^{2}}{6 M_{\rm pl}^{2} H^{2}}\,,\qquad 
\Omega_{I} \equiv \frac{\rho_{I}}{3 M_{\rm pl}^{2} H^{2}}\,,
\qquad 
\tilde{\Omega}_c \equiv \frac{\tilde{\rho}_c}{3M_{\rm pl}^2 H^2}
=\left( 1+Qf \right) \Omega_c\,,
\ee
where $I = r, b, c$.
{}From Eq.~(\ref{Eq00}), the CDM density parameter 
$\tilde{\Omega}_c$, which accommodates the interaction
with the vector field, can be expressed as 
\be
\tilde{\Omega}_c=1-\Omega_{\rm DE}-\Omega_b-\Omega_r\,.
\label{Eq00d}
\ee

Taking the derivatives of $\Omega_{\rm DE}$, $\Omega_{b}$, 
$\Omega_{r}$, and $\Omega_{c}$
with respect to $\mathcal{N} \equiv \ln a$ and using the continuity 
Eqs.~(\ref{rhobeq})-(\ref{rhoceq}), it follows that 
\ba
& &
\Omega_{\rm DE}' = 2 \Omega_{\rm DE} \left(
\epsilon_{\phi} - \epsilon_{h} \right)\,,\label{Omedeeq}\\
& &
\Omega_{b}' = -\Omega_{b} \left(
3+2\epsilon_{h} \right)\,,\label{Omebeq}\\
&&
\Omega_{r}' = -2 \Omega_{r} (2 + \epsilon_{h}) \,, 
\label{Omereq}\\
&& 
\Omega_{c}'= -\Omega_{c} \left(
3+2\epsilon_{h} \right)\,,
\label{Omeceq}
\ea
where 
\be
\epsilon_{\phi} \equiv \frac{\dot{\phi}}{H \phi}\,,\qquad 
\epsilon_{h} \equiv \frac{\dot{H}}{H^2}\,.
\ee
After differentiating Eq.~(\ref{EqA0d}) with respect to $t$ and using 
Eq.~(\ref{Eq11}), we can solve them for $\dot{\phi}$ and $\dot{H}$.
In doing so, we exploit Eqs.~(\ref{Eq00d}) and (\ref{EqA0d}) to 
eliminate $\rho_c$ and $b_3$.
Defining the dimensionless variable
\be
r_{Q} \equiv \frac{Qf_{,X} \rho_c}{b_2}
=-2p_3 Q \left( \frac{u^2}{2} \right)^{2p_3}
\frac{\Omega_{c}}{\Omega_{\rm DE}}\,,\quad 
{\rm with} \quad u \equiv \frac{\phi}{M_{\rm pl}}\,,
\label{r_def}
\ee
we obtain
\ba
\epsilon_{\phi} &=& 
\frac{u'}{u}=
\frac{3(1+r_Q)+(1-r_Q)(\Omega_r-3\Omega_{\rm DE})}
{2(1+r_Q)+2(1-r_Q)^2 s\,\Omega_{\rm DE}}s\,,\label{epphi}\\
\epsilon_{h} &=& \frac{H'}{H}=
-\frac{(1+r_Q)(3+\Omega_r-3\Omega_{\rm DE})
-6r_Q (1-r_Q)s\,\Omega_{\rm DE}}
{2(1+r_Q)+2(1-r_Q)^2s\,\Omega_{\rm DE}}\,,\label{eph}
\ea
where 
\be
s \equiv \frac{1}{2p_3-1}\,.
\ee
In what follows, we focus on the theories with $p_3>1/2$, 
i.e., $s>0$. We also consider the case in which $\Omega_{\rm DE}$ 
is positive, i.e., $b_2<0$.

When $Q=0$, the parameter $s$ characterizes the deviation of 
$w_{\rm DE}$ from $-1$ \cite{DeFelice:2016yws,DeFelice:2016uil}. 
At the background level, our coupled dark energy model has 
two additional parameters $Q$ and $s$ relative to those 
in the $\Lambda$CDM model.
The variable $r_Q$, which corresponds to the ratio between the third and first 
terms on the left hand side of Eq.~(\ref{EqA0d}), obeys the differential equation 
\be
r'_{Q} = r_{Q} \left( \frac{2}{s} \epsilon_{\phi} - 3\right)\,.
\label{rQeq}
\ee

For given constants $Q$ and $s$, the background cosmological 
dynamics is known by integrating
Eqs.~(\ref{Omedeeq})-(\ref{Omereq}) and Eq.~(\ref{rQeq}) 
with Eqs.~(\ref{epphi}) and (\ref{eph}). 
In doing so, we need to specify the initial conditions 
of $\Omega_{\rm DE}$, $\Omega_{b}$, $\Omega_{r}$, and $u$ 
at some redshift $z=1/a-1$.
The initial value of $\Omega_c$ is determined by using Eq.~(\ref{Eq00d}) 
and the correspondence
\be
\Omega_c=\left[ 1+Q \left( \frac{u^2}{2} \right)^{2p_3} 
\right]^{-1} \tilde{\Omega}_c\,.
\label{Omecre}
\ee
{}From Eq.~(\ref{r_def}), the initial condition of $r_Q$ is 
known accordingly. 
Instead of solving Eq.~(\ref{rQeq}), we can also integrate 
Eq.~(\ref{epphi}) for the dimensionless temporal vector 
component $u=\phi/M_{\rm pl}$. 
Nevertheless, using the variable $r_Q$ is convenient to study 
the effect of coupling $Q$ on the background cosmological dynamics. 
Indeed, the dark energy equation of state (\ref{wDEdef}) is
simply expressed as 
\be
w_{\rm DE}=-1-\frac{2}{3}
\left( 1- r_Q \right)\epsilon_{\phi} \,,
\ee
which shows that $w_{\rm DE}$ is determined 
by the two quantities $r_Q$ and $\epsilon_{\phi}$. 

\subsection{Analytic estimation for each cosmological epoch}
\label{anasec}

Before solving the above autonomous system numerically, 
we analytically estimate the evolution of background quantities 
during the radiation, matter, and accelerated epochs.
 
Let us begin with the early Universe in which $\Omega_{\rm DE}$ 
is much smaller than 1. Expanding $\epsilon_{\phi}$ and 
$\epsilon_h$ around $\Omega_{\rm DE}=0$, it follows that 
\be
\epsilon_{\phi}=\frac{3}{2}s+\frac{1-r_Q}{2(1+r_Q)}s \Omega_r
+{\cal O} (\Omega_{\rm DE})\,,\qquad 
\epsilon_h=-\frac{3}{2}-\frac{1}{2}\Omega_r
+{\cal O} (\Omega_{\rm DE})\,.
\label{epest}
\ee
In this regime, the dark energy equation of state can be 
estimated as 
\be
w_{\rm DE} \simeq -1- (1-r_Q)s 
\left[ 1+\frac{1-r_Q}{3(1+r_Q)} \Omega_r 
\right]\,,\quad {\rm for} \quad \Omega_{\rm DE} \ll 1\,.
\label{wedana}
\ee
In addition, the quantity $r_Q$ approximately obeys 
\be
r_Q' \simeq \frac{r_Q(1-r_Q)}{1+r_Q} \Omega_r\,.
\label{rQeq2}
\ee

We would like to consider the case in which the effect of coupling $Q$ 
on Eq.~(\ref{EqA0d}) is unimportant in the early radiation era, 
so that $r_Q \ll 1$. 
{}From the above estimation, the fixed point corresponding to 
the radiation dominance is given by 
\be
{\rm P}_r:\left( \Omega_{\rm DE}, \Omega_b, \Omega_r, 
r_Q \right)=(0,0,1,0)\,,
\ee
with $\tilde{\Omega}_c=0$ and $u=0$. 
In this epoch, the dark energy equation of state (\ref{wedana}) 
reduces to 
\be
w_{\rm DE} \simeq -1-\frac{4}{3}s\,.
\label{wdera}
\ee
This value of $w_{\rm DE}$ is identical to that derived for $Q=0$ 
in Ref.~\cite{DeFelice:2016yws,DeFelice:2016uil}, so 
the effect of coupling $Q$ on $w_{\rm DE}$ 
does not appear in the deep radiation epoch.
However, we have $r_Q' \simeq r_Q$ around the point ${\rm P}_r$
and hence the nonvanishing coupling $Q$ leads to the increase of 
$r_Q$ proportional to $a$.
In the late radiation era, this growth of $r_Q$, 
together with the decrease of $\Omega_{r}$, gives rise to the 
departure of $w_{\rm DE}$ from the constant value (\ref{wdera}).
{}From Eq.~(\ref{epphi}) the temporal vector component obeys 
$u' \simeq 2su$ around the point ${\rm P}_r$, so it grows as 
$u \propto a^{2s}$.

After $\Omega_r$ becomes much smaller than 1 during the matter era, 
we have $\epsilon_{\phi} \simeq 3s/2$ and $\epsilon_h \simeq -3/2$ from 
Eq.~(\ref{epest}). On using Eqs.~(\ref{Omebeq}), (\ref{Omeceq}), and 
(\ref{rQeq}), it follows that $\Omega_b=\Omega_b^{(m)}={\rm constant}$,  
$\Omega_c=\Omega_c^{(m)}={\rm constant}$, and 
$r_Q=r_Q^{(m)}={\rm constant}$ in this epoch. 
Then, the fixed point corresponding to the matter dominance is 
\be
{\rm P}_m:\left( \Omega_{\rm DE}, \Omega_b, \Omega_r, 
r_Q \right)=\left( 0,\Omega_b^{(m)},0,r_Q^{(m)} \right)\,,
\ee
with $\tilde{\Omega}_c=1-\Omega_b^{(m)}$ and $u=0$.
{}From Eq.~(\ref{epphi}), the quantity $u$ approximately obeys 
$u' \simeq 3su/2$ around the point ${\rm P}_m$, so that the 
temporal vector component grows as $u \propto a^{3s/2}$. 
Provided that $|Q(u^2/2)^{2p_3}| \ll 1$, $\Omega_c$ is 
approximately equivalent to $\tilde{\Omega}_c$.
Around the fixed point ${\rm P}_m$, the dark energy equation 
of state is simply given by  
\be
w_{\rm DE} \simeq w_{\rm DE}^{(m)} \equiv -1-\left( 1-r_Q^{(m)} \right) s\,,
\label{wdema}
\ee
where $w_{\rm DE}^{(m)}$ is a constant.
In the limit that $Q \to 0$, this result coincides with the value 
$w_{\rm DE}^{(m)}=-1-s$ derived in Refs.~\cite{DeFelice:2016yws,DeFelice:2016uil}. 
For $Q>0$, $r_Q^{(m)}$ is a negative constant and hence 
$w_{\rm DE}^{(m)}<-1-s$. On the other hand, the negative coupling $Q$ 
gives rise to the value $0<r_Q^{(m)}<1$, so $w_{\rm DE}^{(m)}$ gets closer to $-1$ 
in comparison to the $Q=0$ case.

The magnitude of constant $r_Q^{(m)}$ depends on the initial condition of $r_Q$ 
in the radiation era. In terms of the quantity $Qf$ equivalent to
the second term in the square bracket of Eq.~(\ref{Omecre}), 
we can express $r_Q$ as $r_Q=-2p_3(\Omega_c/\Omega_{\rm DE})Qf$. 
Even if $|Qf|$ is very much smaller than 1, the quantity $r_Q$ can be 
as large as the order of 0.1 due to the large term 
$\Omega_c/\Omega_{\rm DE} \gg 1$ in the matter era.
In Sec.~\ref{nusec}, we will show that the no ghost condition (\ref{qc}) 
puts upper limits on the value $r_Q^{(m)}$.

The matter fixed point ${\rm P}_m$ is different from the 
$\varphi$MDE \cite{Amendola99}
characterized by the nonvanishing constant $\Omega_{\rm DE}$ 
proportional to the coupling-constant squared $\beta^2$, 
in that $\Omega_{\rm DE}=0$ on the point ${\rm P}_m$.
{}From Eq.~(\ref{Omedeeq}), the dark energy density parameter around 
${\rm P}_m$ approximately obeys $\Omega_{\rm DE}'=3(1+s)\Omega_{\rm DE}$, 
so $\Omega_{\rm DE}$ grows as 
\be
\Omega_{\rm DE} \propto a^{3(1+s)}\,.
\ee
Eventually, the contribution of $\Omega_{\rm DE}$ to Eq.~(\ref{rQeq}) 
becomes nonnegligible at the late cosmological epoch. 
Assuming that $r_Q \ll 1$ in this epoch and expanding Eq.~(\ref{rQeq}) 
around $r_Q=0$, it follows that 
\be
r_Q'= -\frac{3(1+s) \Omega_{\rm DE}}
{1+s \Omega_{\rm DE}}r_Q+{\cal O} (r_Q^2)\,.
\ee
This means that, after the dominance of dark energy, 
$r_Q$ starts to decrease from the value $r_Q^{(m)}$. 
Finally, the solutions approach the dS fixed point 
characterized by 
\be
{\rm P}_{\rm dS}:\left( \Omega_{\rm DE}, \Omega_b, \Omega_r, 
r_Q \right)=\left( 1,0,0,0 \right)\,,
\ee
with $\tilde{\Omega}_c=0$ and 
$u=u_{\rm dS}={\rm constant}$. 
Since $\epsilon_{\phi}=0=\epsilon_h$ on this point, 
both $u$ and $H$ are constants, with the dark energy equation 
of state 
\be
w_{\rm DE}=-1\,.
\label{wdeds}
\ee
During the cosmological sequence of radiation, matter, and dS epochs, 
$w_{\rm DE}$ changes as (\ref{wdera}) $\to$ (\ref{wdema}) $\to$ (\ref{wdeds}).

\subsection{Numerical analysis}
\label{nusec}

To study the background cosmological dynamics in more detail, we numerically 
integrate Eqs.~(\ref{Omedeeq})-(\ref{Omereq}) and Eq.~(\ref{rQeq}) for 
several different values of $s$ and $Q$. 
We also discuss whether the conditions for the absence of ghosts and Laplacian 
instabilities for linear cosmological perturbations are consistently satisfied.
The background initial conditions in the deep radiation era 
(around the redshift $z=10^7$) are chosen to realize today's values 
$\Omega_{\rm DE}^{(0)}=0.68$, $\Omega_b^{(0)} \simeq 0.05$, and 
$\Omega_r^{(0)} \simeq 10^{-4}$. We consider the case in which 
today's temporal vector component $u^{(0)}$ is of order 1.

\begin{figure}[h]
\includegraphics[height=3.2in,width=3.3in]{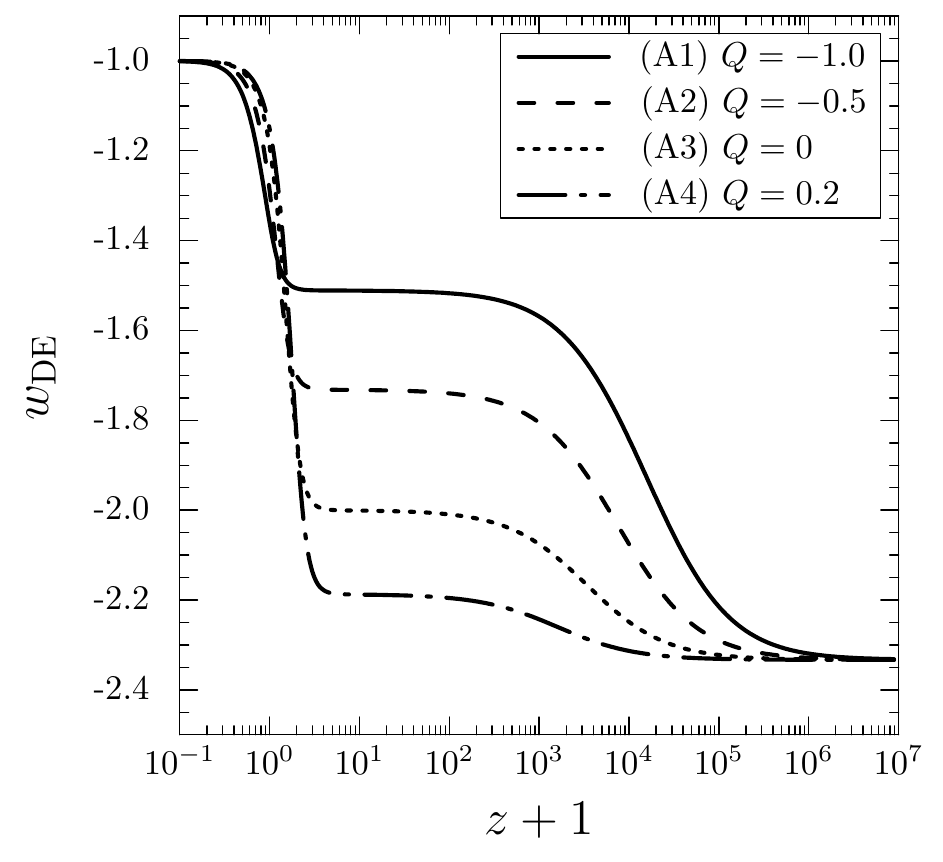}
\includegraphics[height=3.2in,width=3.4in]{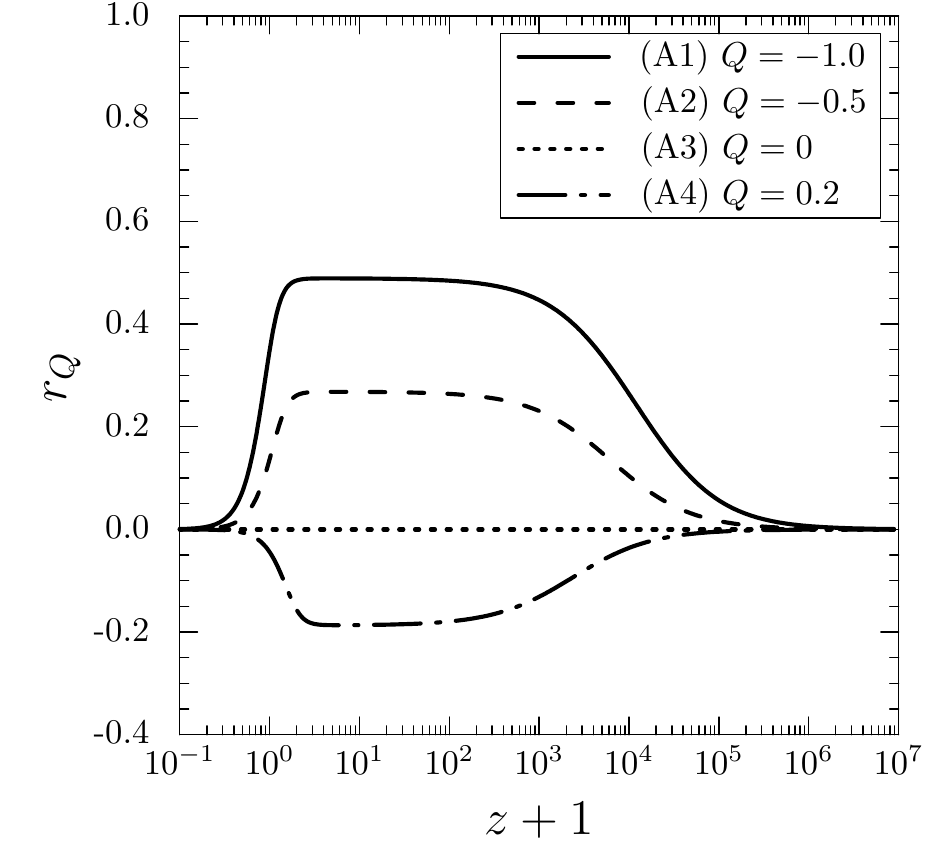}
\caption{\label{fig1}
Evolution of $w_{\rm DE}$ (left) and $r_{Q}$ (right) versus 
$z+1$ for $s=1$ (i.e., $q=2p_3=2$). 
Each line corresponds to the couplings 
(A1) $Q = -1.0$, 
(A2) $Q = -0.5$, 
(A3) $Q =0$, and 
(A4) $Q =0.2$. 
The initial conditions are chosen to give rise to 
today's values $\Omega_{\rm DE}^{(0)}=0.68$, 
$\Omega_b^{(0)} \simeq 0.05$, $\Omega_r^{(0)} \simeq 10^{-4}$, 
and $u^{(0)}=1.04$ (at the redshift $z=0$).
In case (A1), the no-ghost condition $q_{c}>0$ 
is marginally satisfied on the dS fixed point.
}
\end{figure}

In Fig.~\ref{fig1}, we plot the evolution of $w_{\rm DE}$ and 
$r_Q$ versus $z+1$ for $s=1$ and four different values of $Q$. 
When $Q=0$, the dark energy equation of state evolves as 
$w_{\rm DE} = -7/3 \to -2 \to -1$ during the 
radiation, matter, and dS epochs, respectively.
The joint observational constraint based on SN Ia, CMB, BAOs, 
RSDs, $H_0$, and ISW-galaxy cross-correlation data give the 
bound $s=0.185^{+0.100}_{-0.089}$ (95\,\% CL) \cite{Nakamura:2018oyy}, 
so the model with $s=1$ and $Q=0$ is excluded 
due to the large deviation of $w_{\rm DE}$ from $-1$ 
before the onset of cosmic acceleration.

If $Q>0$, we observe in Fig.~\ref{fig1} that $w_{\rm DE}$ is smaller 
than $-2$ during the matter era. 
This is consistent with the analytic estimation (\ref{wdema}), which 
gives $w_{\rm DE}^{(m)}=-2+r_Q^{(m)}<-2$ for $s=1$ and $Q>0$.
In case (A4) of Fig.~\ref{fig1}, the numerical value of $r_Q^{(m)}$ 
is about $-0.2$ and hence $w_{\rm DE}^{(m)} \simeq -2.2$.
This behavior of $w_{\rm DE}$, which occurs through the decay 
of dark energy to CDM, is in more tension with the observational 
data in comparison to the $Q=0$ case. 

When $Q<0$, the quantity $r_Q^{(m)}$ is positive.
In cases (A1) and (A2) of Fig.~\ref{fig1}, we can confirm  
that $r_Q$ temporally approaches the positive constant $r_Q^{(m)}$ 
after its increase during the radiation dominance ($r_Q \propto a$).
This leads to $w_{\rm DE}$ larger than $-2$ in the matter era, 
whose behavior is attributed to the decay of CDM to dark energy.
After the matter-dominated epoch ends, 
$w_{\rm DE}$ starts to approach the asymptotic value $-1$ with 
the decrease of $r_Q$.
The case (A1) in Fig.~\ref{fig1} corresponds to the marginal 
one in which the quantity $q_{c}=1+Qf$ is close to $+0$ at the dS point. 
For $Q<0$, the no-ghost condition $q_c>0$ constrains
the field value $u_{\rm dS}=\phi_{\rm dS}/M_{\rm pl}$ in the range
\be
|Q|\left( \frac{u_{\rm dS}^2}{2} \right)^{2p_3}<1\,.
\label{Qcon}
\ee
This also puts the upper limit on the magnitude of $r_Q^{(m)}$.
When $s=1$, we numerically find the  bound $r_Q^{(m)}<0.48$, 
which translates to $w_{\rm DE}^{(m)}<-1.52$.
Thus, the negative coupling $Q$ allows the 
possibility for realizing $w_{\rm DE}^{(m)}$ closer to $-1$ 
relative to the $Q=0$ case.

\begin{figure}[h]
\includegraphics[height=3.2in,width=3.4in]{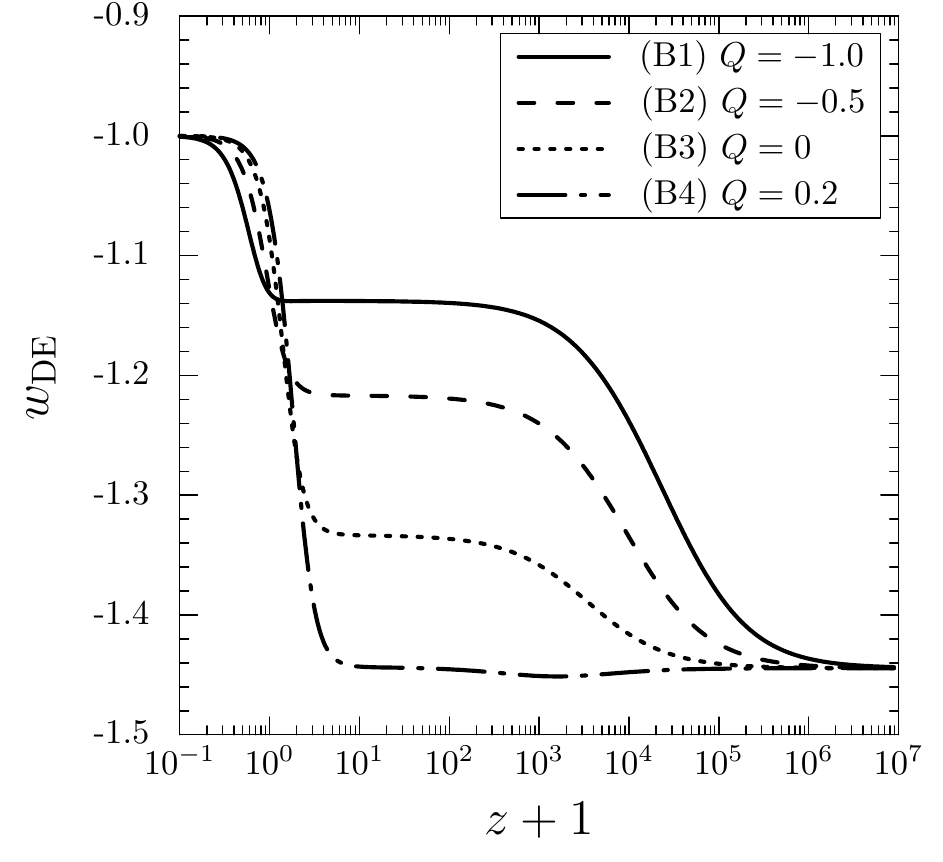}
\includegraphics[height=3.2in,width=3.4in]{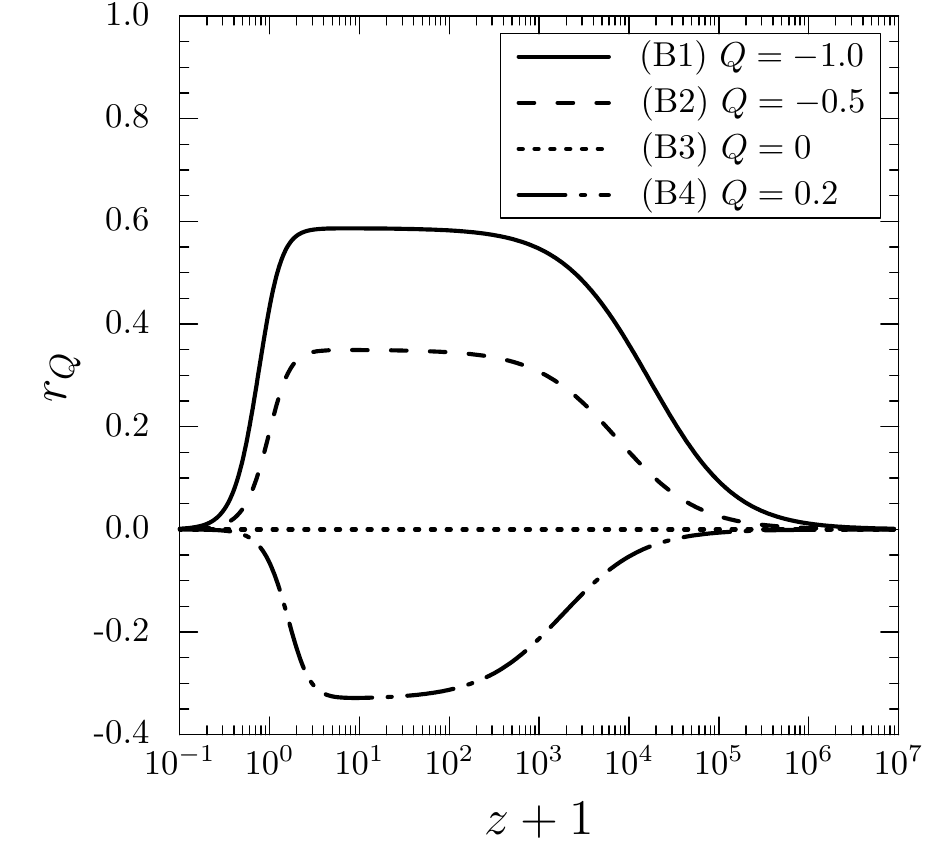}
\caption{\label{fig2}
Evolution of $w_{\rm DE}$ (left) 
and $r_{Q}$ (right) versus $z+1$ 
for $s=1/3$ (i.e, $q=2p_{3}=4$) 
with the different couplings:  
(B1) $Q = -1.0$, 
(B2) $Q = -0.5$, 
(B3) $Q =0$, 
and (B4) $Q = 0.2$. 
Today's values of $\Omega_{\rm DE}$, $\Omega_b^{(0)}$, 
and $\Omega_r^{(0)}$ are the same as those 
in Fig.~\ref{fig1}, while $u^{(0)}=1.166$. 
}
\end{figure}

Although the model with $s=1$ should be still difficult to 
be compatible with the observational data due to the upper bound 
$w_{\rm DE}^{(m)}<-1.52$, the situation 
is different for the models with $s<1$ (i.e., $p_3>1$).
In Fig.~\ref{fig2}, we depict the evolution of $w_{\rm DE}$ and $r_{Q}$ 
for $s=1/3$ (i.e., $p_{3} = 2$) and four different values of $Q$. 
When $Q=0$, we have $w_{\rm DE}^{(m)}=-1.33$, 
in which case the model is outside the 95 \% CL 
observational boundary \cite{Nakamura:2018oyy}. 
As we observe in Fig.~\ref{fig2}, the negative coupling $Q$ leads to 
larger values of $w_{\rm DE}^{(m)}$ compatible with the data.
In case (B2), the numerical values of $w_{\rm DE}$ and $r_{Q}$ 
in the matter era are $-1.22$ and $0.35$, respectively, 
which are consistent with the analytic estimation (\ref{wdema}). 
The case (B1) is the marginal situation in which the quantity 
$q_c$ at the dS point is close to $+0$. 
This corresponds to the maximum dark energy equation 
of state $w_{\rm DE}^{(m)}=-1.14$ with $r_Q^{(m)}=0.59$.
Thus, for $s=1/3$, the negative coupling 
$Q$ gives rise to $w_{\rm DE}^{(m)}$ in the 
range $-1.33<w_{\rm DE}^{(m)}<-1.14$.
This alleviates the problem of observational incompatibility of 
the model with $s=1/3$ and $Q=0$. The positive 
coupling does not improve the situation, see case (B4) 
of Fig.~\ref{fig2}.

\begin{figure}
\includegraphics[width=3.5in]{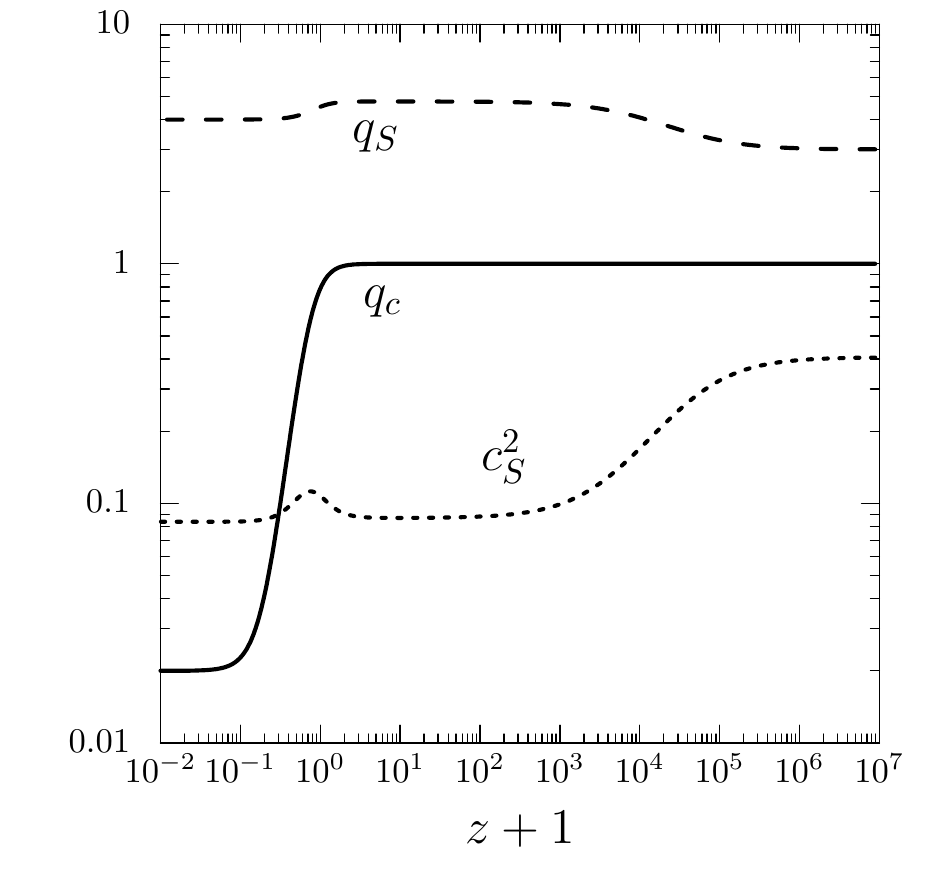}
\caption{\label{fig3}
Evolution of $q_{c}$, $q_{S}$ and $c_{S}^{2}$ 
versus $z+1$ for the case (B1) in Fig.~\ref{fig2}, i.e.,
$s=1/3$ and $Q =-1.0$. 
Today's values of $\Omega_{\rm DE}$, $\Omega_b$, $\Omega_r$, 
and $u$ are chosen in the same 
way as those in Fig.~\ref{fig2}.
The quantity $q_{S}$ is divided by the 
positive quantity 
$12 M_{\rm pl}^{4} H^{2} \Omega_{\rm DE}$.
}
\end{figure}

In the above discussion, the no-ghost condition $q_c>0$ on the dS solution 
is crucial to put an upper bound on the value of $w_{\rm DE}^{(m)}$.
As we showed in Sec.~\ref{anasec}, the temporal vector component 
$u=\phi/M_{\rm pl}$ increases during the radiation and matter epochs.
Around the dS fixed point there is the approximate relation 
$u'/u \simeq 3(1-\Omega_{\rm DE})s/[2(1+s\Omega_{\rm DE})]$, so 
$u$ also grows toward the constant value $u_{\rm dS}$. 
This means that, under the condition (\ref{Qcon}), 
the no-ghost condition $q_c>0$ of CDM is satisfied during 
the whole cosmic expansion history.
The other no-ghost condition (\ref{qS}) of the vector field 
translates to 
\be
q_S=12M_{\rm pl}^4 H^2 \Omega_{\rm DE} 
\left[ \frac{1+r_Q}{s}+(1-r_Q)^2 \Omega_{\rm DE} \right]>0\,.
\label{qsre}
\ee
Provided that $r_Q>-1$, the condition (\ref{qsre}) is always
satisfied. This is the case for $Q<0$, under which $r_Q>0$.

On using the background equations of motion, the vector 
propagation speed squared (\ref{cS}) can be expressed as
\be
c_{S}^{2} = 
\frac{
s \tilde{s} ( 2 s - \tilde{s}) (5 + 3 s)
-2 \tilde{s}^4 \Omega_{\rm DE}^{2}
- \tilde{s}^{2} [ (7 + 3 s) s - 2 (1 + s) \tilde{s} ] \Omega_{\rm DE}
+ s \tilde{s}^{2} (1 + s) \Omega_{r}
}{6 (\tilde{s}^{2} \Omega_{\rm DE} + 2 s -\tilde{s})^{2}} 
+ \frac{2 \tilde{s}^{2} \Omega_{\rm DE}}{3 q_{V} u^{2}
(\tilde{s}^{2} \Omega_{\rm DE} +2 s - \tilde{s})}\,,
\label{csge}
\ee
where $\tilde{s}=(1-r_Q)s$, and $q_V=1$ for the model 
under consideration. 
On the fixed points P$_{r}$, P$_{m}$, and P$_{\rm dS}$, 
Eq.~(\ref{csge}) reduces, respectively, to 
\be
(c_{S}^{2})_{r}=\frac{s (2 s + 3)}{3} \,, \qquad
(c_{S}^{2})_{m}=\frac{s(5+3s)(1-r_Q^{(m)})}
{6(1+r_Q^{(m)})}\,, \qquad
(c_{S}^{2})_{\rm dS}=\frac{2 s}{3 (1 + s) q_{V} 
u_{\rm dS}^{2}}\,.
\label{cS_estimations}
\ee
Since we are considering the case $s>0$, both $(c_{S}^{2})_{r}$ 
and $(c_{S}^{2})_{\rm dS}$ are positive 
under the absence of vector ghosts ($q_V>0$).
During the matter era, the condition $(c_{S}^{2})_{m} \geq 0$ gives 
\be
-1<r_Q^{(m)} \le 1\,,
\label{rQcon}
\ee
which is satisfied for all the cases shown in 
Figs.~\ref{fig1} and \ref{fig2}. 
The quantity $|r_Q|$ reaches the maximum value $|r_Q^{(m)}|$ 
during the matter era, so the no-ghost condition (\ref{qsre}) automatically 
holds under the bound (\ref{rQcon}).

In order to confirm the above analytic estimations, we 
compute the quantities $q_{c}$, $q_{S}$ and $c_{S}^{2}$ by numerically 
integrating the autonomous equations.
Figure \ref{fig3} is such an example, 
which corresponds to the case (B1) in Fig.~\ref{fig2}. 
We recall that this is close to the marginal case in which the condition 
$q_c>0$ is satisfied on the dS solution. 
In Fig.~\ref{fig3}, we observe that $q_c$ starts to deviate from 
1 at low redshifts and it finally approaches the asymptotic 
value $0.02$. As estimated from Eq.~(\ref{qsre}), $q_S$ is always 
positive during the whole cosmological evolution.
Since $s=1/3$, $q_V=1$, $r_Q^{(m)}=0.59$, and $u_{\rm dS}=1.41$ 
for the case (B1) in Fig.~\ref{fig2}, the analytic estimations 
(\ref{cS_estimations}) give the values $(c_{S}^{2})_{r} = 0.407$,
$(c_{S}^{2})_{m} = 0.086$, and $(c_{S}^{2})_{\rm dS}=0.084$, respectively.
They are in good agreement with their numerical values
computed in Fig.~\ref{fig3}. We note that $c_S^2$ also remains 
positive during the transition from the matter era to 
the dS epoch.

As long as the no-ghost condition $q_c>0$ of CDM is satisfied on the dS solution, 
the numerical simulations of Figs.~\ref{fig1} and \ref{fig2} show that 
$|r_Q|$ does not exceed 1. In this case, there are neither ghosts 
($q_S>0$) nor Laplacian instabilities ($c_S^2>0$) for 
the longitudinal scalar mode of $A_{\mu}$.
In other words, under the condition (\ref{Qcon}), all the stability 
conditions associated with scalar perturbations are consistently 
satisfied. We found that the maximum 
allowed values of $w_{\rm DE}$  
consistent with the stability conditions
are $w_{\rm DE}^{(m)}=-1.52$ for $s=1$ and $w_{\rm DE}^{(m)}=-1.14$
for $s=1/3$. For smaller $s$, $w_{\rm DE}^{(m)}$ increases further, 
e.g., $w_{\rm DE}^{(m)}=-1.07$ for $s=1/5$ (i.e., $p_3=3$). 
This upper limit of $w_{\rm DE}^{(m)}$ is mostly determined by 
the product $Qf=Q(u^2/2)^{2p_3}$, which means that both 
coupling constant $Q$ and temporal vector 
component $u$ affect the evolution of $w_{\rm DE}$ 
after the onset of matter dominance.

\section{Conclusions}
\label{concludesec}

We studied the cosmology of GP theories in which a massive 
vector field $A_{\mu}$ is coupled to CDM with the interacting 
action (\ref{Acoupling2}). 
We deal with the matter sector as perfect fluids described by 
the Schutz-Sorkin action (\ref{Schutz}). 
In this approach, the conserved part of the CDM energy 
density $\rho_c$ is associated with the Schutz-Sorkin action, 
while the additional interacting energy density $Qf \rho_c$ 
arises from the interaction (\ref{Acoupling2}).  
As a result, the total effective CDM energy density 
$\tilde{\rho}_c=(1+Qf)\rho_c$ contains the effect of 
coupling $Q$ on the right hand side of Eq.~(\ref{rhoccon}).
Defining the dark energy density and pressure arising from 
the vector field according to Eqs.~(\ref{rhode}) and (\ref{Pde}), 
they obey the modified continuity Eq.~(\ref{rhodecon}), 
whose sign of the interacting term is opposite to 
that of $\tilde{\rho}_c$. This clearly shows the consistency 
of our approach of dealing with the coupling between 
two dark sectors.

In Sec.~\ref{stasec}, we provided a general framework for studying the 
dynamics of linear cosmological perturbations in coupled vector 
dark energy theories given by the action (\ref{GPaction}).
We derived the second-order actions of tensor, vector, and scalar 
perturbations and studied conditions for the absence of ghosts 
and Laplacian instabilities.
The quadratic tensor action is of the same form as that in general 
relativity, so the theories automatically pass the bound on 
the propagation speed of gravitational waves. 
Deep inside the Hubble radius, the second-order action of vector 
perturbations reduces to the form (\ref{SV2}) with 
$q_V=G_{2,F}$ and $c_V^2=1$, so the vector ghost is 
absent under the condition $G_{2,F}>0$. 
For scalar perturbations, there are two no-ghost conditions 
(\ref{qS}) and (\ref{qc}) associated with the vector field and 
CDM, respectively. The scalar propagation speed squared 
$c_S^2$ of the vector field is affected by the coupling $Q$ 
as Eq.~(\ref{cS}), which must be positive 
to avoid the Laplacian instability. We also showed that 
the Laplacian instability is absent in the matter sector.

In Sec.~\ref{modelsec}, we proposed a viable model of coupled 
vector dark energy given by the functions (\ref{G23}). 
For the power $q=2p_3$, the coupling term in Eq.~(\ref{phieq}) 
grows in proportion to $t^{1/2}$ in the radiation era and 
it reaches a constant value during the matter dominance. 
This interacting term starts to decrease after the onset of cosmic 
acceleration, which is followed by the approach to  
de Sitter solutions with $\rho_c=0$.
In other words, the effect of interactions between the vector 
field and CDM on cosmological observables 
mostly manifests itself from the onset of matter era 
to today. During the matter dominance, the dark energy equation of 
state $w_{\rm DE}$ is a constant smaller than $-1$, so the 
corresponding density parameter $\Omega_{\rm DE}$ grows 
in time. This property is different from the $\varphi$MDE 
of coupled scalar dark energy models in which 
$\Omega_{\rm DE}$ is a nonvanishing constant 
affected by the coupling $\beta$. 

We found that the negative coupling $Q$ leads to $w_{\rm DE}$ 
closer to $-1$ relative to the uncoupled case. 
This is attributed to the fact that, for $Q<0$, CDM decays 
to the vector field. The maximum value of 
$w_{\rm DE}$ reached during the matter era is determined by 
the CDM no-ghost condition (\ref{qc}).
In this case, the other stability conditions (\ref{qS}) and (\ref{cS}) 
of scalar perturbations are satisfied from the radiation era to 
the de Sitter attractor.
For $p_3=1, 2, 3$ the maximum values are 
given by $w_{\rm DE}^{(m)}=-1.52, -1.14, -1.07$, respectively, 
which are larger than their corresponding values 
$w_{\rm DE}=-2, -1.33, -1.2$ for $Q=0$. 
Thus, our coupled dark energy model alleviates the 
problem of observational incompatibility of uncoupled 
models with $p_3 \le 2$.

We thus showed that the coupled vector dark energy allows 
the phantom dark energy equation of state being compatible 
with the observational data, while satisfying all the stability 
conditions of linear cosmological perturbations.
This property is very different from the standard coupled quintessence 
in which $w_{\rm DE}=1$ during the $\varphi$MDE. 
It will be of interest to place observational constraints on 
the coupling and the power $p_3$ along the line of 
Ref.~\cite{deFelice:2017paw}. 
On scale relevant to the linear growth of cosmological perturbations, 
the effective gravitational coupling for baryons is different from the Newtown 
gravitational constant due to the existence of cubic interactions 
$G_{3}(X)\nabla_{\mu}A^{\mu}$ \cite{Nakamura:2018oyy}. 
The gravitational interaction for CDM should be further modified by the 
direct coupling to the vector field. 
The derivation of effective gravitational couplings felt by CDM and light is 
the next important step for probing the signatures of coupled vector dark energy 
in the observations of RSDs and ISW-galaxy cross-correlations. 
We leave these interesting issues for future works. 

\section*{Acknowledgements}

RK is supported by the Grant-in-Aid for Young Scientists B 
of the JSPS No.\,17K14297. 
ST is supported by the Grant-in-Aid for Scientific Research Fund of the JSPS No.\,19K03854 and
MEXT KAKENHI Grant-in-Aid for Scientific Research on Innovative Areas
``Cosmic Acceleration'' (No.\,15H05890).


\end{document}